\documentclass[aps,prc,preprint,groupedaddress,showpacs]{revtex4}
\usepackage{amsmath}
\usepackage{amssymb}
\usepackage{bm}
\newcommand{\be}{\begin{equation}}
\newcommand{\ee}{\end{equation}}
\newcommand{\bea}{\begin{eqnarray}}
\newcommand{\eea}{\end{eqnarray}}
\newcommand{\bffr}{\mbox{\boldmath $e$}}

\newcommand{\bfsigma}{\mbox{\boldmath $\sigma$}}
\newcommand{\dbss}[1]{_{\scriptstyle #1}}
\newcommand{\linda}[1]{\vphantom{A^{B}} #1}
\newcommand{\lindb}[1]{\vphantom{A^{B^c}} #1}
\newcommand{\lindc}[1]{\vphantom{A^{B^j}} #1}
\newcommand{\lindd}[1]{\vphantom{A^{b^C}} #1}
\newcommand{\mbss}[1]{_{\mbox{\scriptsize #1}}}

\newcommand{\dsp}{\displaystyle}

\newcommand{\txts}{\textstyle}
\newcommand{\ve}{\varepsilon}
\newcommand{\vkp}{\varkappa}
\newcommand{\vphi}{\varphi}
\newcommand{\vtht}{\vartheta}

\newcommand{\barst}{\bar{S^{\vphantom{1}}}}
\newcommand{\vpb}{\vphantom{\beta}}

\newcommand{\vps}{\vphantom{S}}

\begin{document}
%
%
\title{%
 Majorana spinors and
 extended Lorentz symmetry in four-dimensional theory
 }%
\author{V. I. Tselyaev}
\affiliation{%
 Nuclear Physics Department,
 V. A. Fock Institute of Physics,
 St. Petersburg State University, 198504,
 St. Petersburg, Russia}
\email[E-mail address:  ]{tselyaev@nuclpc1.phys.spbu.ru}
\date{\today}
\begin{abstract}

An extended local Lorentz symmetry in four-dimensional (4D)
theory is considered. A source of this symmetry is
a group of general linear transformations of four-component
Majorana spinors $GL(4,M)$ which is isomorphic to $GL(4,R)$
and is the covering of an extended Lorentz group in a 6D Minkowski
space $M(3,3)$ including superluminal and scaling transformations.
Physical space-time is assumed to be a 4D pseudo-Riemannian manifold.
To connect the extended Lorentz symmetry in the $M(3,3)$ space
with the physical space-time, a fiber bundle over the 4D manifold
is introduced with $M(3,3)$ as a typical fiber.
The action is constructed which is invariant with respect to
both general 4D coordinate and local $GL(4,M)$ spinor transformations.
The components of the metric on the 6D fiber are expressed in terms of
the 4D pseudo-Riemannian metric and two extra complex fields:
4D vector and scalar ones.
These extra fields describe in the general case massive particles
interacting with an extra $U(1)$ gauge field and
weakly interacting with ordinary particles, i.e. possessing properties
of invisible (dark) matter.
\end{abstract}
%

\maketitle

\section{Introduction}

Possible modification of the usual Lorentz symmetry
(from the standpoint of its extension or, on the contrary,
breakdown) is of obvious interest since it underlies
many modern physical theories.
Perhaps, one of the most intriguing questions is an extension
of the Lorentz symmetry admitting superluminal transformations.
These transformations can be introduced immediately in
a flat two-dimensional (2D) space-time (see \cite{P69}).
In this case the generalized Lorentz transformation
between two frames with space-time coordinates $x,t$ and $x',t'$
has the form (in units where $c=1$):
\be
x' = (x - v t)/\sqrt{|1-v^2|}\,,\qquad
t' = (t - v x)/\sqrt{|1-v^2|}
\label{intr1}
\ee
where $v$ is the relative velocity of the frames.
The absolute value of the quadratic form $t^2 - x^2$ is invariant
under this transformation at any $v^2 \neq 1$ but its sign is changed
at $v^2 > 1$ (superluminal case). In the limit $v^2 \to \infty\,$
one obtains: $x' \to \pm t$, $\;t' \to \pm x$.
However, if one starts with the 4D Minkowski space $M(3,1)$,
situation becomes more complicated.
Usually, in this case two ways are considered.
In the first approach, imaginary quantities are introduced
in the 4D or 3D space-time in addition to or instead of
the usual real coordinates
(see Refs. \cite{RM74,D75,MR76,MR84,C77} and references therein
for more details).
In the second approach, the 4D Minkowski space is replaced
by the 6D one $M(3,3)$ by introducing two extra time-like coordinates,
but all the space-time coordinates remain real.
In the $M(3,3)$ space, the superluminal transformations are
the combinations of the 6D rotations, reflections, and
the space-time permutation of the form
\be
x_1 \rightleftarrows \,t_1\,,\qquad
x_2 \rightleftarrows \,t_2\,,\qquad
x_3 \rightleftarrows \,t_3\,.
\label{intr2}
\ee
This approach was proposed in Refs. \cite{D75,MR76},
developed in Refs. \cite{C77,C78,C80,CB82}
(see also references therein), and further elaborated and applied
in Refs. \cite{BY97,BPY97,B98}.
However, both these methods face a question how to reconcile
additional coordinate variables (imaginary or real ones)
with the observed four-dimensionality of the physical space-time.
Possible ways to overcome this difficulty immediately within the
framework of the above-mentioned approaches were discussed in the
papers cited above (see also their discussion and criticism
in Refs. \cite{S80,S81,MAE83}).
In the present paper, the following solution of this problem
is proposed: physical space-time is regarded as
a 4D pseudo-Riemannian manifold which is a base of a vector bundle
with the $M(3,3)$ space as a typical fiber.
The superluminal transformations concern only the fiber of
the frame bundle constructed from this vector bundle.

Another question arising in these approaches is how to introduce
spinor fields in the theory with an extended Lorentz symmetry.
In Refs.~\cite{PS85,BC93}, a spinor representation of the extended
Lorentz group in the $M(3,3)$ space was introduced with the help of
eight-component Dirac spinors.
Corresponding Dirac equations were considered and their properties
and properties of their solutions were analyzed.
However, there remains a question of the interpretation of
additional spinor components in this representation.
As a matter of fact, there is no need to complicate the situation.
The problem is simplified if we choose as a starting point
not the properties of the space-time transformations but
the properties of the spinors. More specifically, we will consider
a group of general linear transformations of four-component
Majorana spinors. This group is the covering
of an extended Lorentz group in the $M(3,3)$ space which includes
superluminal and scaling transformations.
Thus, the extension of the usual Lorentz symmetry is introduced
in a natural way.
On the other hand, Majorana spinors are very simple objects
which can be used to construct usual four-component chiral spinors
without introducing any additional components.

In the present paper,
the action which includes Majorana spinor fields and obeys the extended
symmetry conditions is considered. It will be shown that
its spinor part reduces to the spinor action of the standard 4D theory
under certain gauge conditions. Nevertheless, the action obtained
in this approach contains additional bosonic fields which do not enter
the standard theory.
These fields describe in the general case massive particles possessing,
under certain conditions, the properties of the invisible (dark) matter.
Nowadays, there are numerous arguments in favour of the existence
of the dark matter (DM). At the same time, its structure is unknown
so far, though there are variety of models predicting its existence
and particle content.
A review of the current status of the DM problem can be
found, e.g., in Refs.~\cite{BHS05,OW04}.
Here we consider only the simplest possibility to apply
the present approach to this problem.

The paper is organized as follows.
In Sec.~\ref{elg6d} an outline of the extended Lorentz group
in the $M(3,3)$ space is given.
The group of general linear transformations of four-component
Majorana spinors is described in Sec.~\ref{gl4m}.
In Sec.~\ref{elsphys}
the connection between the extended Lorentz symmetry
and the physical space-time is considered.
The vector bundle and the frame bundle over the 4D manifold
are introduced.
In Sec.~\ref{gsym} the local gauge transformations are considered
and the spin connections are determined.
The chiral spinor fields with the extended
symmetry are introduced in Sec.~\ref{chir}.
In Sec.~\ref{consp} the spinor action is constructed and
the discrete symmetries are considered.
The extra fields arising in the theory and their interpretation
as the dark matter candidates are discussed in Sec.~\ref{extf}.
A summary is given in the last Section.
Appendices contain notations, conventions, auxiliary formulas,
and proof of some equations used in the paper.

\section{Extended Lorentz group
 in six-dimensional space \label{elg6d}}

We start with a brief description of an extended Lorentz group $EL(3,3)$
which is defined in this paper as a group of all linear transformations
of the type $V'^K = \Lambda^K_L\,V^L$. In this equation,
$V^L$ is a vector in the 6D Minkowski space $M(3,3)$
with a metric $\eta_{\dbss{KL}}$ defined in Eq.~(\ref{dan1})
(here and in the following we use notations and conventions
drawn in Appendix~\ref{nots})
and $\Lambda^K_L$ is a real matrix satisfying condition
\be
\Lambda^{K'}_{\lindb{K}}\,\eta_{\dbss{K'L'}}\,
\Lambda^{L'}_{\lindb{L}} =
\vkp\,\eta_{\dbss{KL}},\quad \vkp \neq 0\,.
\label{elg1}
\ee
In the case $\vkp = \pm 1$, these transformations were analyzed
in Refs.~\cite{C78,C80,CB82}. Here we will consider the general case
$\vkp \neq 0$ and supplement the analysis of \cite{C78,C80,CB82}
with some details. Let us represent the matrices
$\Lambda^K_{K'}$ and $\eta_{\dbss{KL}}$ in the form
\be
\Lambda = \left( \begin{array}{cc}
\Lambda^s_{s'} & \Lambda^s_{t'} \\
\Lambda^t_{s'} & \Lambda^t_{t'}
\end{array} \right), \quad
\eta = \left( \begin{array}{cc}
-I_3 & 0 \\ 0 & I_3
\end{array} \right),
\label{elg2}
\ee
where blocks $\Lambda^s_{s'}$, $\Lambda^s_{t'}$,
$\Lambda^t_{s'}$, and $\Lambda^t_{t'}$ are $3 \times 3$ matrices
and $I_3$ is the $3 \times 3$ identity matrix.
Let us first consider two cases.

(i) Subluminal transformations without scaling: $\vkp = +1$.
These transformations form the group $L(3,3)=O(3,3)$ for which
Eq.~(\ref{elg1}) reduces to $\Lambda^T\eta\,\Lambda = \eta$.
(Here and throughout the paper $\Lambda^T$ denotes
a transposed matrix.)
One can show (see Appendix~\ref{detprp})
that in this case the following relations hold:
\be
|\det(\Lambda^s_{s'})| = |\det(\Lambda^t_{t'})|\,
\geqslant \,1\,,
\label{elg3}
\ee
\be
\det(\Lambda) = \det(\Lambda^s_{s'})/\det(\Lambda^t_{t'})\,.
\label{elg4}
\ee
Analogously to the case of the usual 4D Lorentz group, this means
that the group $L(3,3)$ consists of the following four components:
\be
\left.
\begin{array}{lll}
L^{\uparrow}_+(3,3)\,: &   \det(\Lambda) = +1\,, &
\det(\Lambda^t_{t'}) \geqslant +1\,,\\
L^{\uparrow}_-(3,3)\,: &   \det(\Lambda) = -1\,, &
\det(\Lambda^t_{t'}) \geqslant +1\,,\\
L^{\downarrow}_+(3,3)\,: & \det(\Lambda) = +1\,, &
\det(\Lambda^t_{t'}) \leqslant -1\,,\\
L^{\downarrow}_-(3,3)\,: & \det(\Lambda) = -1\,, &
\det(\Lambda^t_{t'}) \leqslant -1\,,\\
\end{array}
\right\}
\label{elg5}
\ee
where $L^{\uparrow}_+(3,3)$ is the proper orthochronous Lorentz
group. Let us introduce two subgroups of the group $L(3,3)$, namely:
$R_{\dbss{ST}} = \{ I_6,\, {\cal{PT}} \}$ and
$R_{\dbss{S}} = \{ I_6,\, {\cal{P}} \}$ where
$I_6$ is the $6 \times 6$ identity matrix,
the matrix ${\cal{PT}}=-I_6$ represents space-time reflection,
and ${\cal{P}}$ is a matrix of the space reflection:
\be
{\cal{P}}^K_{\linda{L}} = \mbox{diag}\{-1,-1,-1,+1,+1,+1\}\,.
\label{elg6}
\ee
The space reflection does not commute with arbitrary
transformation $\Lambda \in L^{\uparrow}_+(3,3)$.
But ${\cal{P}}\,\Lambda \,{\cal{P}}^{-1} \in L^{\uparrow}_+(3,3)$
if $\Lambda \in L^{\uparrow}_+(3,3)$.
From this it follows that the group $L(3,3)$
can be represented as a semidirect product:
$L(3,3) = L^{\uparrow}_+(3,3) \rtimes
(R_{\dbss{S}} \times R_{\dbss{ST}})$.

(ii) Superluminal transformations without scaling: $\vkp = -1$.
Let us introduce subgroup
$P_{\dbss{ST}} = \{ I_6,\, \bar{I}_6 \}$
where $\bar{I}_6$ is a matrix of the space-time permutation:
\be
{\renewcommand{\arraystretch}{0.7}
\bar{I}_6 = \left( \begin{array}{cccccc}
0 & 0 & 0 & 0 & 0 & 1 \\
0 & 0 & 0 & 0 & 1 & 0 \\
0 & 0 & 0 & 1 & 0 & 0 \\
0 & 0 & 1 & 0 & 0 & 0 \\
0 & 1 & 0 & 0 & 0 & 0 \\
1 & 0 & 0 & 0 & 0 & 0
\end{array} \right)\,.
}
\label{elg7}
\ee
We have: $\bar{I}^{\,2}_6 = I^{\vphantom{2}}_6$,
$\,\bar{I}^{\,T}_6\,\eta\,\bar{I}^{\vphantom{2}}_6 = - \eta$,
therefore $P_{\dbss{ST}}$ is a group and
$P_{\dbss{ST}} \subset EL(3,3)$.
Further, if the equality
$\bar{\Lambda}^T\,\eta\,\bar{\Lambda} = -\eta$
holds for some real matrix $\bar{\Lambda}$ and if
$\Lambda = \bar{I}_6\,\bar{\Lambda}$ or
$\Lambda = \bar{\Lambda}\,\bar{I}_6$ then we obtain:
$\Lambda^T\,\eta\,\Lambda = \eta$.
Consequently, any superluminal transformation without scaling
can be represented in the form:
$\bar{\Lambda} = \bar{I}_6\,\Lambda$ or
$\bar{\Lambda} = \Lambda'\,\bar{I}_6$ where
$\Lambda, \Lambda' \in L(3,3)$.
Notice that the equality $[\,\bar{I}_6,\Lambda\,]=0$
does not hold for an arbitrary $\Lambda \in L(3,3)$.
However,
$\bar{I}_6\,\Lambda \,\bar{I}_6{\vphantom{I}}^{-1} \in L(3,3)$
if $\Lambda \in L(3,3)$.

Let us return to the general case and introduce
a group $R^+_{\linda{6}}$ of the scaling transformations in the 6D space
represented by all real matrices of the form
$e^{\lambda} I_6$, $\,\lambda \in \mathbb{R}^1$.
Then from the above analysis it follows that
the extended Lorentz group can be represented as
a semidirect product of the groups $L(3,3)$ and
$P^+_{\linda{ST}} = P_{\lindc{ST}} \times R^+_{\lindb{6}}$, i.e.
$EL(3,3) = L(3,3) \rtimes P^+_{\linda{ST}}\,$.

\section{Group of general linear transformations
 of four-component Majorana spinors \label{gl4m}}

We will see below, and this
is of principle for the present approach, that the covering
of the extended proper orthochronous Lorentz group is a group of
general linear transformations of four-component Majorana spinors.
These spinors are defined by the condition
(see, e.g., \cite{W95}):
\be
\vphi = i \gamma^2 \vphi^*
\label{ms1}
\ee
where $\vphi^*$ stands for the complex conjugate spinor.
From Eq.~(\ref{ms1}) it follows that four-component Majorana spinors
have only two independent complex components. The group of their
general linear non-degenerated transformations
[we will refer to this group as $GL(4,M)$]
is represented by the $4 \times 4$ complex matrices $S$ satisfying
the following conditions:
\be
\gamma^2\, S^{\,*}\, \gamma^2 = -S\,,\quad \det(S) \ne 0\,.
\label{ms2}
\ee
Obviously, if $S \in GL(4,M)$, $\vphi$ satisfies Eq.~(\ref{ms1}),
and $\vphi' = S\vphi$ then $\vphi'$ also satisfies Eq.~(\ref{ms1}).
The following two properties of the group $GL(4,M)$ are important.
\begin{itemize}
\item[(i)]
$SL(2,C) \subset GL(4,M)$.
Thus, the covering of the usual 4D Lorentz group is contained
in $GL(4,M)$.
\item[(ii)]
Let us introduce the matrix
\be
U = \frac{\dsp 1}{\dsp 2\sqrt{2}} \left[ (1-\gamma^5)(1+i\gamma^2)
-i (1+\gamma^5)\,\gamma^4\,(1-i\gamma^2) \right]\,.
\label{ms3}
\ee
It is not difficult to check that $U^{\dag}U = 1$ and
$U^*\gamma^2 = -i\,U$ where $U^{\dag}=(U^*)^T$.
Using these properties we obtain
that if $R = U S \, U^{\dag}$ where
$S \in GL(4,M)$ then $R^* = R$ and $\det(R) \ne 0$.
The opposite is also true: if $S = U^{\dag} R \, U$,
$\,R^* = R$, and $\det(R) \ne 0$ then $S \in GL(4,M)$.
Consequently, $GL(4,M)$ is isomorphic to $GL(4,R)$.
In addition, $\det(S)$ is manifestly real for all
$S$ in $GL(4,M)$.
\end{itemize}

Let us denote:
\be
\barst = \gamma^4\,S^{\,\dag}\,\gamma^4
\label{ms4}
\ee
where $S \in GL(4,M)$. The matrices $\barst$ and
$\barst\vphantom{S}^{-1}$ belong to the conjugate and
conjugate-contragredient (c.-c.) representations of $GL(4,M)$,
respectively. The Majorana spinors of the c.-c. representation
will be denoted by $\chi$. Notice that if
$S \in SL(2,C) \subset GL(4,M)$
then $\barst\vphantom{S}^{-1} = S$ but in the general case
$\barst\vphantom{S}^{-1} \neq S$.
Thus, we have the following four types of the Majorana spinors:
$\vphi$, $\,\bar{\vphi}=\vphi^{\dag}\gamma^4$,
$\,\chi$, and $\bar{\chi}=\chi^{\dag}\gamma^4$, which
transform according to the four different representations
of the group $GL(4,M)$, namely:
\bea
&&\vphi' = S\,\vphi\,,\quad \bar{\vphi}' = \bar{\vphi}\,\barst\,,
\label{tran1}\\
&&\chi' = \barst\vphantom{S}^{-1}\chi\,,\quad
\bar{\chi}' = \bar{\chi}\,S^{-1}\,.
\label{tran2}
\eea

Consider the quantities
$\bar{\vphi}_{\lindc{1}}\,\alpha^K \vphi_{\lindd{2}}$ and
$\bar{\chi}_{\lindc{1}}\,\tilde{\alpha}_{\dbss{K}} \chi_{\lindd{2}}$
where matrices $\alpha^K$ and $\tilde{\alpha}_{\dbss{K}}$
are determined by Eqs. (\ref{mprp1}) and (\ref{mprp2})
(see Appendices \ref{nots} and \ref{matprp} for the
definitions and properties of these and other matrices
introduced in the paper).
As follows from Eqs. (\ref{tran1}) and (\ref{tran2}),
the transformation laws of these quantities are determined
by the matrix products $\barst\,\alpha^K S$ and
$S^{-1} \tilde{\alpha}_{\dbss{K}} \barst\vphantom{S}^{-1}$.
In Appendix~\ref{prfeqs} it is proved that if $S \in GL(4,M)$
then
\bea
&&\barst\,\alpha^K S = \Lambda^K_{\lindb{L}} \alpha^{L}\,,
\label{ms5}\\
&&S^{-1} \tilde{\alpha}_{\dbss{K}} \barst\vphantom{S}^{-1} =
\tilde{\Lambda}^{L}_{\lindb{K}}\,\tilde{\alpha}_{\dbss{L}}\,,
\label{ms8}
\eea
where
\bea
&&\Lambda^K_{\lindb{L}} = \frac{1}{4}\,
\mbox{Tr}\left( \barst\,\alpha^K S\,
\tilde{\alpha}_{\dbss{L}} \right)
= \Lambda^{K*}_{\lindb{L}}\,,
\label{ms6}\\
&&\tilde{\Lambda}^{L}_{\lindb{K}} = \frac{1}{4}\,
\mbox{Tr}\left( S^{-1} \tilde{\alpha}_{\dbss{K}}
\barst\vphantom{S}^{-1} \alpha^{L}\right)
= \tilde{\Lambda}^{L\,*}_{\lindb{K}}\,.
\label{ms9}
\eea
In Appendix~\ref{prfeqs} it is also proved that the matrices
$\Lambda$ and $\tilde{\Lambda}$ defined by Eqs.
(\ref{ms6}) and (\ref{ms9}) form a group,
that $\tilde{\Lambda} = \Lambda^{-1}$,
and that the following relations hold:
\bea
&&\Lambda^{K'}_{\lindb{K}}\,\eta_{\dbss{K'L'}}\,
\Lambda^{L'}_{\lindb{L}} = \det(S)\,\eta_{\dbss{KL}}\,,
\label{ms13}\\
&&\tilde{\Lambda}^{K'}_{\lindb{K}}\,\eta_{\dbss{K'L'}}\,
\tilde{\Lambda}^{L'}_{\lindb{L}} =
\det(S^{-1})\,\eta_{\dbss{KL}}\,.
\label{ms21}
\eea
Eqs. (\ref{ms13}) and (\ref{ms21}) mean that the matrices
$\Lambda$ and $\tilde{\Lambda}$ belong to the group $EL(3,3)$
described in Sec.~\ref{elg6d}.
Thus, one can conclude that the quantities
$\bar{\vphi}_{\lindc{1}}\,\alpha^K \vphi_{\lindd{2}}$ and
$\bar{\chi}_{\lindc{1}}\,\tilde{\alpha}_{\dbss{K}} \chi_{\lindd{2}}$
transform under spinor $GL(4,M)$ transformations
as a vector and a covector in the $M(3,3)$ space,
respectively.

Let us determine the relation between the groups $GL(4,M)$ and
$EL(3,3)$ more precisely.
The mapping $GL(4,M) \to EL(3,3)$ is determined by Eqs.
(\ref{ms6}) or (\ref{ms9}).
Further,
the matrices $S \in GL(4,M)$ with $\det(S)=+1$ form a subgroup
$SL(4,M)$ which is isomorphic to $SL(4,R)$. Since the group
$SL(4,R)$ is connected
and the mappings (\ref{ms6}) and (\ref{ms9}) are continuous,
the group $SL(4,M)$ is mapped only to the one connected component
of the group $EL(3,3)$, namely {\it onto} the proper orthochronous
Lorentz group $L^{\uparrow}_+(3,3)$
(see the remark at the end of Appendix~\ref{prfeqs}).
As a result, by including space-time permutation and scaling
transformations (see Appendix~\ref{prfeqs})
we obtain that $GL(4,M)$ is the covering of the group
$L^{\uparrow}_+(3,3) \rtimes P^+_{\linda{ST}}$
which is a subgroup of $EL(3,3)$.
The correspondence between the total extended Lorentz group
$EL(3,3)$ and the spinor transformations will be considered
in Sec.~\ref{consp}.

\section{Extended Lorentz symmetry
 and the physical space-time \label{elsphys}}

Now our aim is to connect the extended Lorentz symmetry
in the $M(3,3)$ space with the physical space-time.
We adopt the traditional model and regard the latter
as a 4D pseudo-Riemannian manifold ${\cal{M}}_{\linda{4}}$.
The vectors and tensors on this manifold and its coordinates
$x^{\mu}$ will be labelled by the Greek letters
$\mu,\nu,\ldots \in \{0,1,2,3\}$. In particular,
$g_{\mu\nu}=g_{\mu\nu}(x)$ will be the metric with $-2$ signature.
To introduce the spinor fields with $GL(4,M)$ symmetry on
${\cal{M}}_{\linda{4}}$, we first construct a vector bundle with
${\cal{M}}_{\linda{4}}$ as a base and a 6D vector space
${\cal{F}}_{\linda{6}}$ as a typical fiber
(see, e.g., \cite{EGH80} for a review of the theory
of fiber bundles and its physical applications).
The fiber is defined as a direct sum:
${\cal{F}}_{\linda{6}}=T^*_x({\cal{M}}_{\linda{4}})+\mathbb{R}^2$
where $T^*_x({\cal{M}}_{\linda{4}})$ is the cotangent space of
${\cal{M}}_{\linda{4}}$ at a point $x$
and $\mathbb{R}^2$ is the associated plane.
We will suppose that there exists a family of the preferred bases for
${\cal{F}}_{\linda{6}}$ which is constructed in the following way.
The set of the exact one-forms $\bffr^{\mu}(x)=dx^{\mu}$
is taken as a basis for the subspace $T^*_x({\cal{M}}_{\linda{4}})$.
Further, it is supposed that in $\mathbb{R}^2$, a family of pairs of
the linearly independent one-form fields $\bffr^u(x)$ and $\bffr^v(x)$
is given. All the pairs in this family are related
by the following $SO(2)$ transformation:
\be
\bffr^{\prime u} = \cos \vtht\,\bffr^u - \sin \vtht\,\bffr^v\,,\quad
\bffr^{\prime v} = \sin \vtht\,\bffr^u + \cos \vtht\,\bffr^v
\label{rot}
\ee
where $\vtht = \vtht(x)$.
Thus, the family of the preferred bases for ${\cal{F}}_{\linda{6}}$
is represented by the sets consisting of the six fields $\bffr^{\mu}(x)$,
$\bffr^u(x)$, and $\bffr^v(x)$, or, in shorthand notations, by the sets
$\{\bffr^F(x)\}$, $F \in \{0,1,2,3,u,v\}$.
It is supposed that the theory should be invariant under general coordinate
transformations $x \to x'(x)$ and the local $SO(2)$ transformation
(\ref{rot}) in $\mathbb{R}^2$.

Let $\mbox{\boldmath $G$}$ be the metric tensor on
${\cal{F}}_{\linda{6}}$.
In terms of the preferred basis $\{\bffr^F\}$, the metric
$\mbox{\boldmath $G$}$ has the components
$G^{FF'}=\mbox{\boldmath $G$}(\bffr^F,\bffr^{F'})$.
It is supposed that $\mbox{\boldmath $G$}$ has the signature
of the metric of the $M(3,3)$ space.
So, we can always find an orthonormal basis
in which the metric $\mbox{\boldmath $G$}$ is proportional to
$\eta^{KK'}$ [see Eq.~(\ref{fmet2}) below].
Therefore, the vector space ${\cal{F}}_{\linda{6}}$
can be identified with $M(3,3)$.
However, in the general case, the matrix $G^{FF'}$ cannot be taken
in the diagonal or block-diagonal form because two one-forms
$\bffr^F$ and $\bffr^{F'}$ entering the preferred basis
are not generally orthogonal with respect to $\mbox{\boldmath $G$}$.
In particular, the components $G^{0u}$, $G^{0v}$, and $G^{uv}$
characterize deviation of the time gradient $dx^0$ from the
direction perpendicular to the plane $\mathbb{R}^2$.
The connection of $G^{FF'}$ with the metric $g_{\mu\nu}$ on the
base manifold will be determined below.

An orthonormal basis $\{\mbox{\boldmath $E$}^K(x)\}$
for the ${\cal{F}}_{\linda{6}}$ fiber
at a point $x$ of the base can be constructed with the help of
a set $\{E^K_{\linda{F}}(x)\}$ consisting of the 36 real fields
which constitute the $6 \times 6$ non-singular matrix (local frame).
By definition, we have:
$\mbox{\boldmath $E$}^K = E^K_{\linda{F}}\bffr^F$.
The fields $E^K_{u}$ and $E^K_{v}$ are invariant
under general coordinate transformations.
The field $E^K_{\mu}$ determines a mapping of
the tangent space of ${\cal{M}}_{\linda{4}}$ into $M(3,3)$.
It is assumed that the local frame transforms under
spinor $GL(4,M)$ gauge transformations as
$\,E'^{K}_{\linda{F}} = \Lambda^{K}_{\linda{L}}E^L_{\linda{F}}$
where the matrix $\Lambda^{K}_{\linda{L}} \in EL(3,3)$
is determined by Eq.~(\ref{ms6}). The orthonormality condition for
the basis $\{\mbox{\boldmath $E$}^K\}$ reads
\be
E^K_{\linda{F}}\,G^{FF'} E^{K'}_{\linda{F'}} = \sigma \eta^{KK'}
\label{fmet2}
\ee
where $\sigma$ is a quantity transforming under $GL(4,M)$
according to $\sigma' = \det(S)\,\sigma$.
It is introduced to ensure invariance of the metric $G^{FF'}$
under corresponding transformations of the local frame.
The form of $\sigma$ is specified below in Eq.~(\ref{frm1}).
Inverting Eq.~(\ref{fmet2}), we obtain:
\be
G_{\lindb{FF''}}G^{F''F'}=\delta^{F'}_{\linda{F}}\,,\quad
G_{\lindb{FF'}} = \sigma^{-1}
E^K_{\linda{F}}\,\eta_{\linda{KK'}} E^{K'}_{\linda{F'}}\,.
\label{fmet3}
\ee

The set of all frames $\{E^K_{\linda{F}}(x)\}$
is the fiber at a point $x$
of the frame bundle over ${\cal{M}}_{\linda{4}}$ constructed from
the given vector bundle. Thus, in this scheme the extended Lorentz
symmetry including superluminal transformations concerns only
the fiber of the frame bundle.
It is convenient to combine the two real fields $E^K_{u}$ and $E^K_{v}$
into the one complex field $W^K=E^K_{u}+iE^K_{v}$.
Then the set $\{E^K_{\mu},W^K\}$ will play the role of the
local frame which is the basic set of variables related to the
space-time sector of the theory. Let us denote
\be
\sigma = \frac{1}{2}\,W^*_{\linda{K}}W^K\,,\quad
\tau = (2\sigma)^{-1}W_{\lindb{K}}W^{K}
\label{frm1}
\ee
(we remind that the 6D indices are lowered and raised by the
constant matrices $\eta_{\dbss{KL}}$ and $\eta^{KL}$).
According to Eq.~(\ref{ms13}), the quantity $\sigma$ in
Eq.~(\ref{frm1}) transforms under the $GL(4,M)$ transformations as
$\sigma' = \det(S)\,\sigma$, while $\tau$ is invariant.
Notice that Eq.~(\ref{frm1}) for $\sigma$ reduces the number of
the independent components of the metric $G^{FF'}$ from 21 to 20,
since Eqs. (\ref{fmet3}) and (\ref{frm1}) lead to the following
constraint: $G_{\linda{uu}}+G_{\linda{vv}}=2$.

Let us now define
\bea
e^K_{\mu} &=& E^K_{\mu} -
\frac{1}{2}\,\bigl(f^*_{\mu}z^K + f_{\linda{\mu}}z^{K*}
\bigr)\,,
\label{frm3a}\\
z^K &=& \sqrt{\frac{\xi - 1}{2\,\xi}}\,
\bigl(\,\xi\,W^K - h\,W^{K*} \bigr)
\label{frm3b}
\eea
where
\be
f_{\linda{\mu}} = \sigma^{-1}z_{\lindb{K}}\,E^K_{\mu}\,,\quad
h = \tau/\sqrt{1 - \tau^*\tau}\,,\quad
\xi = 1 + \sqrt{1 + h^*h}\,.
\label{frm3c}
\ee
It is easy to verify that the 6D vectors $e^K_{\mu}$ and $z^K$
satisfy the following conditions:
\be
z_{\dbss{K}}e^K_{\mu} = 0\,,\quad
z_{\dbss{K}}z^K = 0\,,\quad
z^*_{\lindb{K}}z^K = 2\,\sigma\,.
\label{frm4}
\ee
Eqs. (\ref{frm1})--(\ref{frm4}) are valid at any values of the fields
$E^K_{\mu}$ and $W^K$ if $\sigma \neq 0$ and $\tau^*\tau<1$.
In this region, there is a one-to-one correspondence
between the local frame $\{E^K_{\mu},W^K\}$ and the set of the fields
$\{e^K_{\mu},z^K,f_{\linda{\mu}},h,\sigma\}$
constrained by the conditions (\ref{frm4}).
Owing to these constraints, the number of the independent
real variables in both sets is equal to 36.
Thus, the replacement of the frame $\{E^K_{\mu},W^K\}$ by the set
$\{e^K_{\mu},z^K,f_{\linda{\mu}},h,\sigma\}$ used in the following
is the simple change of variables.
On the other hand, the set
$\{e^K_{\mu},z^K,f_{\linda{\mu}},h,\sigma\}$ is constrained
in addition to Eqs.~(\ref{frm4}) only by the conditions $\sigma \neq 0$
and $\det(g_{\lindb{\mu\nu}}) \neq 0$ (see below).
There are no additional constraints on the fields
$f_{\linda{\mu}}$ and $h$. This fact facilitates further analysis
and explains the choice of these variables.

Formally, we have:
$\mbox{\boldmath $E$}^K = E^K_{\linda{F}}\bffr^F
= e^K_{\linda{F}}\bffr^{\,\prime F}$ where
$e^K_u=\mbox{Re}(z^K)$ and $e^K_v=\mbox{Im}(z^K)$.
Thus, the set $\{e^K_{\mu},z^K\}$ is a new local frame which
connects two bases $\{\mbox{\boldmath $E$}^K\}$ and
$\{\bffr^{\,\prime F}\}$, and in which the real and the imaginary
parts of the vector $z^K$ form the orthogonal subset.
Notice, however, that the basis $\{\bffr^{\,\prime F}\}$,
which is transformed into $\{\mbox{\boldmath $E$}^K\}$ with the help
of the frame $\{e^K_{\mu},z^K\}$, is not generally the preferred one.
So, the fields $f_{\linda{\mu}}$, $h$, and $\sigma$ are necessary
to determine all the components of the metric
$\mbox{\boldmath $G$}$ in terms of the preferred basis.

Let us define the relation between
the metric $g_{\lindb{\mu\nu}}$ on ${\cal{M}}_{\linda{4}}$ and the
local frame by the ansatz
\be
g_{\lindb{\mu\nu}} = \sigma^{-1}e_{\lindb{K\mu}}e^K_{\linda{\nu}}\,.
\label{frm2}
\ee
The matrix $g_{\lindb{\mu\nu}}$ is invertible if the matrix
$E^K_{\linda{F}}$ is non-singular and $\sigma \neq 0$.
The inverse matrix
$g^{\mu\nu}$ will be used to raise the 4D coordinate indices.
Using Eqs. (\ref{fmet3})--(\ref{frm2}), one can show after some algebra
that the connection between $g_{\lindb{\mu\nu}}$ and the metric $G^{FF'}$
on the fiber is determined by the natural relation
$G^{\mu\nu}=g^{\mu\nu}$. The other components of $G^{FF'}$ are expressed
in terms of the fields $h$ and $f_{\linda{\mu}}$.
Invertibility of $g_{\lindb{\mu\nu}}$ in combination
with Eqs. (\ref{frm4}) and (\ref{frm2}) leads to the equality
\be
\sigma^{-1}e^K_{\linda{\mu}}e^{\mu}_{\linda{L}} + (2\sigma)^{-1}
\bigl( z^{K*}z_{\lindb{L}} + z^{K}z^*_{\linda{L}}\bigr) =
\delta^K_{\linda{L}}
\label{frm9}
\ee
which is the completeness relation for the set $\{e^K_{\mu},z^K\}$.
Notice that the fields $h$, $f_{\linda{\mu}}$,
and $g_{\lindb{\mu\nu}}$ are invariant under the spinor $GL(4,M)$
transformations and contain 20 independent real components.
Together with 16 parameters of the $GL(4,M)$ group,
this yields 36 degrees of freedom that is equal to the number of
components of the local frame $\{E^K_{\linda{F}}\}$.

It is interesting to note that if we would define the metric
on ${\cal{M}}_{\linda{4}}$ as
$\tilde{g}_{\lindb{\mu\nu}} = e_{\linda{K\mu}}e^K_{\nu}$
we would arrive at the Weyl geometry (see, e.g., Refs. \cite{R82,I99}
where the Weyl theory and its modifications are described in detail).
In the present case the Weyl gauge transformation is generated
by the scaling transformation of the $GL(4,M)$ group.
We will not use this approach because it leads to the known
difficulties. Only the metric (\ref{frm2}), which is invariant
under the $GL(4,M)$ transformations including scaling, enters
the all subsequent formulas.

The set of the fields $\{e^K_{\mu},z^K,f_{\linda{\mu}},h,\sigma\}$
constrained by Eqs.~(\ref{frm4}) can be further reduced to the
following set: $\{e^a_{\mu},f_{\linda{\mu}},h,{\hat\theta}\}$
where $e^a_{\mu}$ is the usual vierbein,
$a \in \{1,2,3,4\}$ (see Appendix~\ref{nots}), and
${\hat\theta}$ is a set consisting of the 10 parameters
of the 16-parameter gauge group $GL(4,M)$.
The set $\{e^a_{\mu},f_{\linda{\mu}},h,{\hat\theta}\}$
is determined in the following way.
It is not difficult to show that if the fields
$e'^K_{\mu}$, $z'^K$, and $\sigma'$
satisfy Eqs.~(\ref{frm4}), they can always be represented in the form
\be
e'^{K}_{\mu} = \Lambda^{K}_{\linda{L}}({\hat\theta})\,e^L_{\mu}\,,
\quad
z'^{K} = \Lambda^{K}_{\linda{L}}({\hat\theta})\,z^L\,,
\quad
\sigma' = \det(S({\hat\theta}))\,\sigma
\label{stg}
\ee
where
\be
\sigma=1\,,\quad
z^a=0\,,\quad z^5=1\,,\quad z^6=i\,,\quad
e^5_{\mu}=e^6_{\mu}=0
\label{csa3}
\ee
and the matrix $\Lambda^{K}_{\linda{L}}({\hat\theta})$ is determined
by Eq.~(\ref{ms6}) with $S=S({\hat\theta}) \in GL(4,M)$.
The 10-parameter set ${\hat\theta}$
is sufficient for this representation. We will assume that
the gauge (\ref{csa3}) corresponds to zero values of the
parameters ${\hat\theta}$. Thus, in the representation (\ref{stg})
and (\ref{csa3}) we again have 36 independent real components
including the vierbein $e^a_{\mu}$,
the fields $f_{\linda{\mu}}$ and $h$, and the set of the
gauge parameters ${\hat\theta}$.

\section{Spinor transformations and local gauge symmetry
 \label{gsym}}

Consider spinor transformations (\ref{tran1}) and (\ref{tran2})
in which the matrix $S$ depends on the point of the space-time
manifold. Let $D_{\mu}=\partial_{\mu}+\Omega_{\mu}$ be
a covariant derivative of the spinor field $\vphi$
where $\Omega_{\mu}$ is the spin connection.
From the condition $D'_{\mu}S\vphi=SD_{\mu}\vphi$ we obtain
the following transformation law:
\be
\Omega'_{\mu} = S\,\Omega_{\mu}S^{-1} - (\partial_{\mu}S)S^{-1}\,.
\label{gcov1}
\ee
Using the complete set $\{1,S_{\lindb{KL}}\}$ (see Appendix~\ref{matprp})
let us represent the spin connection in the form
\be
\Omega^{\vphantom{KL}}_{\mu} =
\frac{i}{2}\,\Omega^{KL}_{\mu}S_{\lindb{KL}}
+ \lambda_{\dbss{\,\mu}}
\label{gcov2}
\ee
where $\Omega^{KL}_{\mu}$ and $\lambda_{\dbss{\,\mu}}$
are the gauge fields related to the matrix $\Omega_{\mu}$
by the formulas [see Eqs. (\ref{mprp22}) and (\ref{mprp25})]
\be
\Omega^{KL}_{\mu} = -i\,\mbox{Tr}\,(\,\Omega_{\mu}S^{KL})\,,
\quad\lambda_{\dbss{\,\mu}} = \frac{1}{4}\,
\mbox{Tr}\,(\,\Omega_{\mu})\,.
\label{gcov3}
\ee
Since both $\vphi$ and $D_{\mu}\vphi$ should be Majorana spinors,
the matrix $\Omega_{\mu}$ satisfies the condition
[cf. Eq.~(\ref{ms2})]
$\gamma^2\,\Omega^*_{\mu}\,\gamma^2=-\Omega^{\vphantom{*}}_{\mu}$.
From this it follows that the fields
$\Omega^{KL}_{\mu}$ and $\lambda_{\dbss{\,\mu}}$ are real.
Further, from the definitions (\ref{ms6}), (\ref{mprp5}) and
Eqs. (\ref{ms5}), (\ref{ms8}), (\ref{ms13}), and (\ref{ms12})
it follows that
\be
S^{-1}S^{KL}S = \det(S^{-1})\,
\Lambda^{K}_{\lindb{K'}}\,\Lambda^{L}_{\lindb{L'}}S^{K'L'}
\label{gcov4}
\ee
where matrices $\Lambda^{K}_{\lindb{K'}}$ are determined by
Eq.~(\ref{ms6}).
Using Eqs. (\ref{ms6})--(\ref{ms13}), (\ref{mprp5}), (\ref{mprp7}),
(\ref{mprp23}), and (\ref{ms12}) one can also obtain
\be
\mbox{Tr}\,(\,S^{-1}S^{KL}\partial_{\dbss{\mu}}S) =
\frac{i}{2}\,\det(S^{-1})\,\bigl(\,
\Lambda^{K}_{\lindb{K'}}\,\partial_{\dbss{\mu}}\,
\Lambda^{L}_{\lindb{L'}}-
\Lambda^{L}_{\lindb{L'}}\,\partial_{\dbss{\mu}}\,
\Lambda^{K}_{\lindb{K'}}
\bigr)\,\eta^{K'L'}\,.
\label{gcov5}
\ee
In addition we have:
\be
\mbox{Tr}\,(\,S^{-1}\partial_{\dbss{\mu}}S) =
\det(S^{-1})\,\partial_{\dbss{\mu}}\det(S)\,.
\label{gcov6}
\ee
As a result,
Eqs. (\ref{gcov1})--(\ref{gcov6}) lead to the following formulas
of the transformations of the gauge fields
$\Omega^{KL}_{\mu}$ and $\lambda_{\dbss{\,\mu}}$:
\be
\Omega'^{KL}_{\mu} = \det(S^{-1})\,\bigl[\,
\Lambda^{K}_{\lindb{K'}}\,\Lambda^{L}_{\lindb{L'}}\,
\Omega^{K'L'}_{\mu} - \frac{1}{2}\bigl(\,
\Lambda^{K}_{\lindb{K'}}\,\partial_{\dbss{\mu}}\,
\Lambda^{L}_{\lindb{L'}}-
\Lambda^{L}_{\lindb{L'}}\,\partial_{\dbss{\mu}}\,
\Lambda^{K}_{\lindb{K'}} \bigr)\,\eta^{K'L'}\bigr]\,,
\label{gcov7}
\ee
\be
\lambda'_{\dbss{\,\mu}} = \lambda_{\dbss{\,\mu}}
- \frac{1}{4}\,\det(S^{-1})\,\partial_{\dbss{\mu}}\det(S)\,.
\label{gcov8}
\ee

We do not regard these gauge fields as independent variables here.
Instead, we express them in terms of the local frame variables
in analogy with the standard approach, which uses the vierbein formalism.
Let us introduce the field
\bea
\omega^{KL}_{\mu} &=& \frac{1}{2\sigma}\,
(\,e^K_{\linda{\nu;\,\mu}}\,e^{L\nu}
-  e^L_{\linda{\nu;\,\mu}}\,e^{K\nu})
\nonumber\\
&-& \frac{1}{4\sigma}\,(\,
z^{K*}\partial_{\mu}z^L + z^K\partial_{\mu}z^{L*} -
z^{L*}\partial_{\mu}z^K - z^L\partial_{\mu}z^{K*} )
\label{gcov9}
\eea
where $e^K_{\linda{\nu;\,\mu}}$ is the covariant derivative
with respect to the conventional metric connection
$\Gamma^{\lambda}_{\mu\nu} = \frac{1}{2}\,
g^{\lambda\kappa}\,(\,\partial_{\linda{\mu}}g_{\linda{\kappa\nu}}
+\partial_{\linda{\nu}}g_{\linda{\mu\kappa}}
-\partial_{\linda{\kappa}}g_{\linda{\mu\nu}})$,
i.e.,
$\,e^K_{\linda{\nu;\,\mu}}=\partial_{\lindb{\mu}}e^K_{\linda{\nu}}
-\Gamma^{\lambda}_{\linda{\mu\nu}}e^K_{\linda{\lambda}}$.
We also put
\be
\lambda_{\dbss{\,\mu}} = - \frac{1}{4}\,
\sigma^{-1}\partial_{\linda{\mu}}\sigma\,.
\label{gcov12}
\ee
Taking into account that the vectors of the local frame
$\{e^K_{\mu},z^K\}$ and the quantity $\sigma$ transform under
spinor $GL(4,M)$ transformations as
$\,e'^K_{\linda{\mu}} = \Lambda^{K}_{\lindb{L}}e^L_{\linda{\mu}}$,
$\,z'^K = \Lambda^{K}_{\lindb{L}}z^L$, and
$\sigma' = \det(S)\,\sigma\,$ and using Eq.~(\ref{frm9}),
we obtain transformations laws (\ref{gcov7}) and (\ref{gcov8})
for the gauge fields determined by
Eqs. (\ref{gcov9}) and (\ref{gcov12}).

Consider the additional $SO(2)$ symmetry determined by Eq.~(\ref{rot}).
Corresponding local $U(1)$ transformation of the fields $W^K$ and $z^K$
reads
\be
W'^K = e^{i\vtht}W^K\,,\quad
z'^K = e^{i\vtht}z^K\,.
\label{gcov13}
\ee
Let us define ($x$-dependent) matrix $\gamma_{\lindb{\bot}}$ as
\be
\gamma_{\lindb{\bot}} = i\,\sigma^{-1}
z^{K*}z^L\,S_{\lindb{KL}}\,.
\label{gcov14}
\ee
Let $C_{\mu}$ be a gauge field and let $C_{\mu}$ is changed
under the transformation (\ref{gcov13}) as
\be
C\,'_{\mu} = C_{\mu} + \partial_{\linda{\mu}}\vtht\,.
\label{gcov15}
\ee
Let us put
\be
\Omega^{\vphantom{KL}}_{\mu} = \omega_{\dbss{\mu}}
+ \lambda_{\dbss{\,\mu}}
+ \frac{i}{2}\,C_{\linda{\mu}} \gamma_{\lindb{\bot}}
\label{gcov16}
\ee
where
\be
\omega_{\lindb{\mu}} =
\frac{i}{2}\,\omega^{KL}_{\mu}S_{\lindb{KL}}
\label{gcov16a}
\ee
and the fields $\omega^{KL}_{\mu}$ and $\lambda_{\dbss{\,\mu}}$
are determined by Eqs. (\ref{gcov9}) and (\ref{gcov12}).
From the above derivation it follows that the spin connection
(\ref{gcov16}) is changed under the local $GL(4,M)$ transformations
according to Eq.~(\ref{gcov1}) and is invariant under the local
$U(1)$ symmetry (\ref{gcov13}).

Spin connection $\tilde{\Omega}_{\mu}$ for the spinor field $\chi$
of the c.-c. representation is determined from the condition
$\tilde{D}'_{\mu}\barst\vphantom{S}^{-1}\chi=
\barst\vphantom{S}^{-1}\tilde{D}_{\mu}\chi\,$ [see Eq.~(\ref{tran2})]
where $\tilde{D}_{\mu}=\partial_{\mu}+\tilde{\Omega}_{\mu}$.
Analogously to the previous case we obtain
\be
\tilde{\Omega}^{\vphantom{KL}}_{\mu} = \tilde{\omega}_{\dbss{\mu}}
- \lambda_{\dbss{\,\mu}}
- \frac{i}{2}\,C_{\linda{\mu}} \tilde{\gamma}_{\lindb{\bot}}
\label{gcov17}
\ee
where
\bea
\tilde{\omega}_{\lindb{\mu}} &=&
\frac{i}{2}\,\omega^{KL}_{\mu}\tilde{S}_{\lindb{KL}}\,,
\label{gcov17a}\\
\tilde{\gamma}_{\lindb{\bot}} &=& -i\,\sigma^{-1}
z^{K*}z^L\,\tilde{S}_{\lindb{KL}}\,.
\label{gcov18}
\eea

\section{Chiral spinor fields \label{chir}}

We now have at our disposal all necessary elements
to construct the spinor action which is invariant under both
local $GL(4,M)$ and general coordinate transformations.
However, there are still two little problems.
First, if a spinor $\vphi$ satisfies constraint (\ref{ms1})
then $i\vphi$ does not satisfy it, and consequently $i\vphi$ is not
a Majorana spinor. In other words, we cannot multiply the
Majorana spinors by the arbitrary complex numbers without
violating the Majorana constraint. As a result, we cannot immediately
introduce internal symmetries of the Majorana spinor fields based,
e.g., on the unitary groups.
Second, to construct a realistic theory one needs chiral
spinor fields. In the standard approach their role
is played by the Weyl spinors.
The four-component Weyl spinors have the same number of the
independent complex components, namely two, as the Majorana spinors,
but they obey different constraint which is not invariant under the
$GL(4,M)$ transformations (see, e.g., \cite{CW83} for
the analysis of the questions related to the chirality
and the properties of the different types of the spinors).

These problems can be solved in the following way. First of all,
notice that from the definitions (\ref{gcov14}) and (\ref{gcov18})
and Eqs. (\ref{frm4}) and (\ref{sprod}) it follows that
\be
\gamma^2_{\linda{\bot}}=\tilde{\gamma}^2_{\linda{\bot}}=1\,.
\label{csa1}
\ee
Therefore, the matrices
\be
\Pi_{\lindb{\pm}} = \frac{1}{2}\,
\bigl(\,1\pm\gamma_{\lindb{\bot}}\bigr)\,,\quad
\tilde{\Pi}_{\lindb{\pm}} = \frac{1}{2}\,
\bigl(\,1\pm\tilde{\gamma}_{\lindb{\bot}}\bigr)
\label{csa2}
\ee
are the projection operators.
In the gauge (\ref{csa3}), using definitions of Appendix~\ref{matprp}
we obtain
\be
\gamma_{\lindb{\bot}}=\tilde{\gamma}_{\lindb{\bot}}=\gamma^5\,,
\label{csa4}
\ee
\be
\Pi_{\lindb{-}}=\tilde{\Pi}_{\lindb{-}}=P_{\lindb{L}}\,,\quad
\Pi_{\lindb{+}}=\tilde{\Pi}_{\lindb{+}}=P_{\lindb{R}}
\label{csa5}
\ee
where $P_{\lindb{L}}$ and $P_{\lindb{R}}$ are the usual chiral
projection operators.

Let us introduce generalized Weyl spinors $\vphi_{\lindb{\pm}}$
and the spinors of the c.-c. representation $\chi_{\lindb{\pm}}$
as follows:
\be
\vphi_{\lindb{\pm}} = \Pi_{\lindb{\pm}}\vphi\,,\quad
\chi_{\lindb{\pm}} = \tilde{\Pi}_{\lindb{\pm}}\chi
\label{csa6}
\ee
where $\vphi$ and $\chi$ are the Majorana spinors.
Obviously, they obey the following generalized Weyl constraint:
\be
\gamma_{\lindb{\bot}}\vphi_{\lindb{\pm}} =
\pm\vphi_{\lindb{\pm}}\,,\quad
\tilde{\gamma}_{\lindb{\bot}}\chi_{\lindb{\pm}} =
\pm\chi_{\lindb{\pm}}\,.
\label{csa7}
\ee
In the gauge (\ref{csa3}) $\vphi_{\lindb{-}}$ and $\chi_{\lindb{-}}$
will be the left-handed spinors and
$\vphi_{\lindb{+}}$ and $\chi_{\lindb{+}}$ will be the right-handed ones.
So, in what follows we will also use notations:
$\vphi_{\lindb{L}}=\vphi_{\lindb{-}}$ and
$\chi_{\lindb{R}}=\chi_{\lindb{+}}$.
Thus, the spinors $\vphi_{\lindb{\pm}}$ and $\chi_{\lindb{\pm}}$
play the role of the chiral spinor fields in the approach under
consideration. Notice that in order to introduce the chiral spinors
within this approach it is of principle to reduce the dimensionality
of the physical space-time to four as compared with the 6D fiber.

The generalized Weyl spinors transform
under the $GL(4,M)$ transformations like the Majorana spinors
[see Eqs. (\ref{tran1}) and (\ref{tran2})], i.e.
\bea
&&\vphi'_{\linda{\pm}} = \Pi'_{\linda{\pm}}\vphi'
= S\,\vphi_{\lindb{\pm}}\,,
\label{csa8}\\
&&\chi'_{\linda{\pm}} = \tilde{\Pi}\vphantom{\Pi}'_{\linda{\pm}}\chi'
= \barst\vphantom{S}^{-1}\chi_{\lindb{\pm}}
\label{csa9}
\eea
where the matrices $\Pi'_{\linda{\pm}}$ and
$\tilde{\Pi}\vphantom{\Pi}'_{\linda{\pm}}$ are defined by
Eqs. (\ref{gcov14}), (\ref{gcov18}), and (\ref{csa2}) in which
$\sigma$ and $z^K$ are replaced by $\sigma' = \det(S)\,\sigma$
and $z'^K = \Lambda^{K}_{\lindb{L}}z^L$. This replacement yields
\be
\gamma'_{\linda{\bot}} = S\,\gamma_{\lindb{\bot}}S^{-1}\,,
\quad
\tilde{\gamma}'_{\linda{\bot}} = \barst\vphantom{S}^{-1}
\tilde{\gamma}_{\lindb{\bot}}\barst\vphantom{S}\,.
\label{csa10}
\ee
From Eqs. (\ref{csa8})--(\ref{csa10}) it follows that the
constraint (\ref{csa7}) is invariant under the $GL(4,M)$
transformations.

Despite the fact that the spinors $\vphi_{\lindb{\pm}}$ and
$\chi_{\lindb{\pm}}$ transform like the Majorana spinors,
they do not satisfy the Majorana constraint (\ref{ms1}).
Instead, we have
\be
\vphi_{\lindb{\pm}} = i \gamma^2 \vphi^*_{\linda{\mp}}\,,\quad
\chi_{\lindb{\pm}} = i \gamma^2 \chi^*_{\linda{\mp}}\,.
\label{csa11}
\ee
An important property of the generalized Weyl spinors is that
in contrast to the Majorana spinors they admit multiplication
by the complex numbers. Indeed, from Eqs. (\ref{csa1}),
(\ref{csa2}), and (\ref{csa6}) it follows that
\be
\vphi'_{\linda{\pm}} = e^{\pm i\theta}\vphi_{\lindb{\pm}}
= \Pi_{\lindb{\pm}}\vphi'
\label{csa12}
\ee
where $\vphi' = S(\theta)\vphi$,
$\,S(\theta)=\exp(i\,\theta\,\gamma_{\lindb{\bot}})$,
and $\theta$ is real.
Using the definition (\ref{gcov14}) one can check that
$S(\theta) \in GL(4,M)$.
Then both $\vphi$ and $\vphi'$ are the Majorana spinors,
and consequently both $\vphi_{\lindb{\pm}}$ and
$\vphi'_{\linda{\pm}}$ in Eq.~(\ref{csa12})
are the generalized Weyl spinors. The same is true for the
spinors $\chi_{\lindb{\pm}}$.
Taking into account this property, we will assume that the spinors
$\vphi_{\lindb{L}}=\vphi_{\lindb{-}}$ and
$\chi_{\lindb{R}}=\chi_{\lindb{+}}$ transform
under the phase transformation (\ref{gcov13}) as follows
\be
\vphi'_{\linda{L}} = e^{\,i\,(q-\frac{1}{2})\,\vtht}\,
\vphi_{\lindb{L}}\,,\quad
\chi'_{\linda{R}} = e^{\,i\,(q-\frac{1}{2})\,\vtht}\,
\chi_{\lindb{R}}
\label{csa13}
\ee
where $q$ is a reduced gauge charge.

\section{Construction of the spinor action
         and the discrete\\ symmetries \label{consp}}

Let us define the following $x$-dependent matrices:
\be
\beta_{\lindb{\mu}} = \frac{i}{2}\,\sigma^{-2}\,
e^K_{\linda{\mu}}z^L z^{M*}\,\beta_{\lindb{KLM}}\,,\quad
\tilde{\beta}_{\lindb{\mu}} = \frac{i}{2}\,\sigma^{-1}\,
e^K_{\linda{\mu}}z^L z^{M*}\,\tilde{\beta}_{\lindb{KLM}}\,.
\label{csa14}
\ee
These matrices are invariant under the simultaneous
$GL(4,M)$ transformations of the type (\ref{ms5}) and (\ref{ms8})
and the $EL(3,3)$ transformations of the local frame.
In the gauge (\ref{csa3}) they take the form
\be
\beta_{\lindb{\mu}} = \tilde{\beta}_{\lindb{\mu}}
= \gamma_{\lindb{\mu}}\,,\quad
\gamma_{\lindb{\mu}} = e^a_{\linda{\mu}} \gamma_{\lindb{a}}\,.
\label{csa15}
\ee

Consider the spinor Lagrangian of the following form
(in units where $\hbar=1$ and $c=1$):
\be
{\cal{L}}_{\mbss{spin}} = \frac{i}{2}\,\bigl(\,
\bar{\vphi}_{\lindb{L}}\,\beta^{\mu}\,
\nabla_{\lindb{\mu}}\vphi_{\lindb{L}}
+ \bar{\chi}_{\lindb{R}}\,\tilde{\beta}\vpb^{\mu}\,
\tilde{\nabla}_{\lindb{\mu}}\chi_{\lindb{R}}\bigr)
- m\,\bar{\chi}_{\lindb{R}}\,\vphi_{\lindb{L}}
+ \,\mbox{H.c.}
\label{csa16}
\ee
where $m$ is a mass parameter,
\be
\nabla_{\lindb{\mu}} = \partial_{\lindb{\mu}} +
\omega_{\lindb{\mu}} - i\,q\,C_{\lindb{\mu}}\,,\quad
\tilde{\nabla}_{\lindb{\mu}} = \partial_{\lindb{\mu}} +
\tilde{\omega}_{\lindb{\mu}} - i\,q\,C_{\lindb{\mu}}\,,
\label{csa17}
\ee
and the spin connections $\omega_{\lindb{\mu}}$ and
$\tilde{\omega}_{\lindb{\mu}}$ are defined by
Eqs. (\ref{gcov16a}) and (\ref{gcov17a}).
The spinor fields $\vphi_{\lindb{L}}$ and $\chi_{\lindb{R}}$
are supposed to be anticommuting.
The covariant derivatives $\nabla_{\linda{\mu}}$ and
$\tilde{\nabla}_{\linda{\mu}}$ differ from the derivatives
$D_{\linda{\mu}}$ and $\tilde{D}_{\linda{\mu}}$ introduced
in Sec.~\ref{gsym} first, by the absence of the gauge field
$\lambda_{\linda{\mu}}$ disappearing due to the hermiticity
of the Lagrangian ${\cal{L}}_{\mbss{spin}}$.
Second, in Eq.~(\ref{csa17}) the properties (\ref{csa7})
of the generalized Weyl spinors were taken into account
and the charges under the $U(1)$ gauge field $C_{\linda{\mu}}$
were changed in order to compensate additional contributions
arising from the phase transformations (\ref{csa13}).

Corresponding spinor action reads
\be
{\cal{S}}_{\mbss{spin}} = \int d^4x\,\sqrt{-g}\,
{\cal{L}}_{\mbss{spin}}
\label{csa18}
\ee
where $g=\det(g_{\lindb{\mu\nu}})$. By construction, this action
is invariant under both general 4D coordinate and local $GL(4,M)$
spinor transformations. In addition it is invariant under the local
$U(1)$ transformations defined by Eqs. (\ref{gcov13}) and
(\ref{csa13}). On the other hand, from Eq.~(\ref{csa15})
it follows that in the gauge (\ref{csa3}) and at $q=0$,
$\,{\cal{S}}_{\mbss{spin}}$ reduces to the action of the
standard 4D theory for the massive neutral fermions without
additional internal symmetries.

Consider the discrete symmetries of the Lagrangian (\ref{csa16}).
We will consider only the symmetries concerning transformations
of the spinors and the vectors in the fiber of the bundle,
since the invariance of the theory with respect to the
general coordinate transformations is implied.
Let us define the following  simultaneous
transformation of the Majorana spinors $\vphi_{\lindb{L}}$ and
$\chi_{\lindb{R}}$ and the vectors of the local frame
$\{E^K_{\mu},W^K\}$:
\bea
&&\vphi'_{\linda{L}} = i\gamma^4\chi_{\lindb{R}}\,,\quad
\chi'_{\linda{R}} = i\gamma^4\vphi_{\lindb{L}}\,,
\label{dsym1}\\
&&E'^K_{\linda{\mu}} = \sigma^{-1}
{\cal{P}}^K_{\linda{L}} E^L_{\linda{\mu}}\,,\quad
W'^K = \sigma^{-1}{\cal{P}}^K_{\linda{L}} W^L
\label{dsym2}
\eea
where the matrix of the space reflection ${\cal{P}}$ and
the quantity $\sigma$ are defined by Eqs. (\ref{elg6}) and
(\ref{frm1}), respectively.
The transformation (\ref{dsym2}) is a combination of the
space reflection and the non-linear conformal transformation
$E'^K_{\linda{\mu}} = \sigma^{-1}E^K_{\linda{\mu}}$,
$W'^K = \sigma^{-1}W^K$. So, in what follows it will be
referred to as $P_{\linda{C}}$ transformation.
The combination of the transformations (\ref{dsym1}) and (\ref{dsym2})
will be referred to as $P$ transformation.
Using definitions (\ref{gcov9}), (\ref{gcov16a}), (\ref{gcov17a}),
and (\ref{csa14}) and the properties (\ref{sref2}) and (\ref{sref3})
one can show that the matrices
$\beta_{\lindb{\mu}}$, $\tilde{\beta}_{\lindb{\mu}}$,
$\omega_{\lindb{\mu}}$, and $\tilde{\omega}_{\lindb{\mu}}$
are changed under the $P_{\linda{C}}$ transformation
as follows:
\bea
&&\beta'_{\linda{\mu}} =
\gamma^4\,\tilde{\beta}_{\lindb{\mu}}\gamma^4\,,\quad
\tilde{\beta}'_{\linda{\mu}} =
\gamma^4\,\beta_{\lindb{\mu}}\gamma^4\,,
\label{dsym3}\\
&&\omega'_{\linda{\mu}} =
\gamma^4\,\tilde{\omega}_{\lindb{\mu}}\gamma^4\,,\quad
\tilde{\omega}'_{\linda{\mu}} =
\gamma^4\,\omega_{\lindb{\mu}}\gamma^4\,.
\label{dsym4}
\eea
Since the fields $\vphi_{\lindb{L}}$ and $\chi_{\lindb{R}}$
enter the Lagrangian (\ref{csa16}) in a symmetric way,
from Eqs. (\ref{dsym3}) and (\ref{dsym4}) it follows that
${\cal{L}}_{\mbss{spin}}$ is invariant under the $P$ transformation.
This invariance can be broken if, for instance,
the fields $\vphi_{\lindb{L}}$ and $\chi_{\lindb{R}}$ possess
different internal symmetries.

The $PT$ transformation is defined as a combination of the
spinor transformation
\be
\vphi'_{\linda{L}} = i\gamma^2\vphi^*_{\linda{L}}\,,\quad
\chi'_{\linda{R}} = -i\gamma^2\chi^*_{\linda{R}}\,,
\label{dsym5}
\ee
the permutation of the spinor fields in each term of
${\cal{L}}_{\mbss{spin}}$, the space-time reflection in the fiber
\be
E'^K_{\linda{\mu}} = - E^K_{\linda{\mu}}\,,\quad
W'^K = - W^K\,,
\label{dsym6}
\ee
and the change of the sign of the field $C_{\lindb{\mu}}$.
From Eqs. (\ref{mprp11}), (\ref{mprp13}), (\ref{mprp14}),
(\ref{mprp19}), and (\ref{mprp20}) it follows that the spinor Lagrangian
is invariant under the $PT$ transformation defined in this way.

The $CPT$ transformation is a combination of the spinor
transformation
\be
\vphi'_{\linda{L}} = \vphi_{\lindb{L}}\,,\quad
\chi'_{\linda{R}} = -\chi_{\lindb{R}}\,,
\label{dsym7}
\ee
the permutation of the spinor fields in ${\cal{L}}_{\mbss{spin}}$,
and the space-time reflection (\ref{dsym6}). Obviously,
the Lagrangian (\ref{csa16}) is invariant under this transformation.

Finally, notice that, as shown in Sec.~\ref{elg6d}, the group $L(3,3)$
can be constructed from the proper orthochronous Lorentz group
$L^{\uparrow}_+(3,3)$ by adding the space and the space-time
reflections. Thus, we arrive at the result:
the group $GL(4,M)$ supplemented by the $P$ and the $PT$
transformations is the covering of the total
extended Lorentz group $EL(3,3)$.

\section{Extra fields as the dark matter candidates
\label{extf}}

Consider now the actions for the other fields introduced
in Sections \ref{elsphys} and \ref{gsym}.
Let us take them in the simplest form.
In particular, for the gravitational action we adopt
the standard Einstein-Hilbert ansatz
\be
{\cal{S}}_{\linda{g}} = -\frac{1}{16 \pi G_{\linda{N}}}
\int d^4x\,\sqrt{-g}\,R
\label{extf1}
\ee
where $R$ is the Ricci scalar curvature
of the space-time with metric $g_{\lindb{\mu\nu}}$:
\be
R = g^{\mu\nu} \bigl(
\partial_{\lindb{\lambda}}\Gamma^{\lambda}_{\linda{\mu\nu}} -
\partial_{\lindb{\nu}}\Gamma^{\lambda}_{\linda{\mu\lambda}} +
\Gamma^{\kappa}_{\linda{\lambda\kappa}} \Gamma^{\lambda}_{\linda{\mu\nu}} -
\Gamma^{\lambda}_{\linda{\mu\kappa}} \Gamma^{\kappa}_{\linda{\lambda\nu}}
\bigr)
\label{sccur}
\ee
and $G_{\linda{N}}$
is the Newtonian constant. There remain three extra fields:
$h$, $f_{\linda{\mu}}$, and $C_{\linda{\mu}}$.
In contrast to the Kaluza-Klein theories (see, e.g., \cite{OW97}),
these fields are not connected with the metric $g_{\lindb{\mu\nu}}$
by any symmetry conditions. So, the Lagrangian for them can be
constructed independently.

Let us define the $U(1)$ covariant derivative of the field
$h$:
\be
h_{\linda{|\mu}} = (\,\partial_{\linda{\mu}} - 2 i C_{\linda{\mu}})
\,h
\label{extf2}
\ee
and the strength tensors of the fields
$f_{\linda{\mu}}$ and $C_{\linda{\mu}}$:
\bea
f_{\linda{[\mu|\nu]}} &=&
(\,\partial_{\linda{\nu}} - iC_{\linda{\nu}})\,f_{\linda{\mu}} -
(\,\partial_{\linda{\mu}} - iC_{\linda{\mu}})\,f_{\linda{\nu}}\,,
\label{extf3}\\
C_{\linda{\mu\nu}} &=& \partial_{\linda{\mu}}C_{\linda{\nu}} -
\partial_{\linda{\nu}}C_{\linda{\mu}}\,.
\label{extf4}
\eea
The simplest form of the action including extra fields is:
\be
{\cal{S}}_{\mbss{ext}} = \int d^4x\,\sqrt{-g}\,
{\cal{L}}_{\mbss{ext}}
\label{extf5}
\ee
where
\be
{\cal{L}}_{\mbss{ext}} =
\frac{1}{16 \pi G_{\linda{N}}}\,\bigl[\,\alpha_{\lindb{h}}
\bigl(\,h^*_{\linda{|\mu}}h^{|\mu} - m^2_{\linda{h}}\,h^*h\,
\bigr) - \alpha_{\lindb{f}}\bigl(\,{\txts{\frac{1}{2}}}\,
f^*_{\linda{[\mu|\nu]}}f^{[\mu|\nu]} - m^2_{\linda{f}}\,
f^*_{\linda{\mu}}f^{\mu}\,\bigr)
\bigr]
- \frac{1}{16 \pi \alpha_{\lindb{C}}}\,
C_{\linda{\mu\nu}}\,C^{\mu\nu}\,.
\label{extf6}
\ee
In Eq.~(\ref{extf6}), $\alpha_{\lindb{h}}$, $\alpha_{\lindb{f}}$,
and $\alpha_{\lindb{C}}$ are the dimensionless constants and
$m_{\lindb{h}}$ and $m_{\lindb{f}}$ are the mass parameters.
${\cal{L}}_{\mbss{ext}}$ is the standard Lagrangian describing
two massive complex bosonic fields (scalar and vector) and one
massless field of the Maxwell type. However, there are no apparent reasons
to identify the field $C_{\linda{\mu}}$ with the usual electromagnetic
field or with the $U(1)_Y$ gauge field of the Standard Model \cite{W96}.
It is more natural to associate the extra fields
$h$, $f_{\linda{\mu}}$, and $C_{\linda{\mu}}$ with the fields
of the invisible (dark) matter.
Notice that these fields do not correspond to tachyons in spite of
the superluminal Lorentz symmetry of the theory.

Consider the interaction of the extra fields with
the ordinary (visible) matter represented here by the fermionic
fields discussed above. This interaction is mediated by the local
frame $\{E^K_{\mu},W^K\}$ and the $U(1)$ gauge field
$C_{\linda{\mu}}$. The fields $h$ and $f_{\linda{\mu}}$
are expressed in terms of the local frame via Eqs. (\ref{frm1}),
(\ref{frm3b}), and (\ref{frm3c}).
On the other hand, the fermionic fields are
connected with the local frame by means of Eqs.
(\ref{frm1})--(\ref{frm3c}), (\ref{frm2}), (\ref{gcov9}),
(\ref{gcov14}), (\ref{gcov16a}), (\ref{gcov17a}), (\ref{gcov18}),
(\ref{csa2}), (\ref{csa6}), (\ref{csa14}), (\ref{csa16}), and
(\ref{csa17}). But in the gauge (\ref{csa3}) this connection
reduces to the dependence of ${\cal{L}}_{\mbss{spin}}$ on the
vierbein $e^a_{\mu}$, while the fields $h$ and $f_{\linda{\mu}}$
become independent variables.
Thus, in this gauge the fermions are connected with the fields
$h$ and $f_{\linda{\mu}}$ only via the metric $g_{\lindb{\mu\nu}}$
(i.e., via gravity) and via the field $C_{\linda{\mu}}$.
The $U(1)$ charges of the fields $h$ and $f_{\linda{\mu}}$
are equal to $2$ and $1$, respectively,
(in units of $\sqrt{\alpha_{\lindb{C}}}$) while the charge of the
fermionic field is equal to $q$. In principle, the value of $q$ is
arbitrary and it can be different for the different kinds of the
fermions. If the value of $q\sqrt{\alpha_{\lindb{C}}}$ is small or
equal to zero, the interaction between the extra fields and the
ordinary matter is weak. It allows us to consider the extra fields
as the dark matter candidates.

There are variety of models predicting the existence of the
non-baryonic DM particles. Among those one should mention
the supersymmetric models and the models with universal extra dimensions
of the Kaluza-Klein type (see \cite{BHS05,OW97,FKM07} for a review),
axion models \cite{OW04,RB00}, and the model of the mirror matter
\cite{F04,B04}. Perhaps there exist several kinds of the fields
corresponding to the DM particles.
Notice that the possibility of the application of the 6D relativity
\cite{C77,C78,C80,CB82} in the $M(3,3)$ space to the DM problem
was discussed in Ref.~\cite{C00}.
In the present paper the same $M(3,3)$ space is introduced.
However, in contrast to Refs.~\cite{C77,C78,C80,CB82,BY97,BPY97,B98},
this space is related here to the fiber of the bundle
but not to the physical space-time.
In particular for this reason, the present treatment
of the consequences of the theory in terms of the DM
hypothesis differs from the approach of Ref.~\cite{C00}.

Let us emphasize that the Lagrangians of the fermionic, gravitational,
and extra fields were taken here in the simplest form.
In particular, the internal symmetries of the fermionic fields
were not taken into account and
the possible couplings of the extra $U(1)$ gauge field
$C_{\linda{\mu}}$ to the gauge fields of these internal symmetries
and to the Higgs bosons were not considered.
The couplings of this and other types were studied in the extensions
of the Standard Model which include an extra $U(1)$ gauge boson
(see \cite{AMP95,L99,ADH03,D05} and references therein).
If these couplings are introduced, the field $C_{\linda{\mu}}$
can acquire a mass due to the spontaneous breaking of the $U(1)$
symmetry. In addition, in principle, it is possible to introduce
direct couplings of the extra fields $h$ and $f_{\linda{\mu}}$
to each other and their Yukawa couplings to the fermionic fields
(of the ordinary or the dark matter) which have different charges
$q$ under the extra $U(1)$ symmetry (see, e.g., discussion
of the similar couplings in Ref.~\cite{BF04} in the context of the
model of the scalar DM). However, the analysis of these questions
is beyond the scope of the present work.

\section{Summary}

A central question of the paper is how to include the extended
superluminal Lorentz symmetry in the four-dimensional (4D) theory.
It is known that this symmetry can be consistently introduced
in the 6D Minkowski space $M(3,3)$.
In this paper it is shown that the problem of reconciling
the 6D relativity with the 4D theory can be solved if it is assumed
that the extended Lorentz group acts only in the fiber of the bundle
for which the physical space-time serves as a base.
The latter is regarded as a 4D pseudo-Riemannian manifold.
In this approach, a source of the extended Lorentz symmetry
is a group of general linear transformations of
four-component Majorana spinors.
The fiber bundle allows us to construct the action including
the spinor fields with the extended symmetry on the 4D manifold.
It is shown that the spinor part of the action reduces to
the spinor action of the standard 4D theory under a certain
choice of the gauge.

The action obtained contains extra fields describing in the
general case two massive particles (4D vector and scalar ones)
weakly interacting with ordinary particles.
The third extra $U(1)$ gauge field is massless.
There are no tachyons in the theory.
If the charges of the fermionic fields under the extra
$U(1)$ symmetry are equal to zero, the extra fields interact
with the ordinary matter only via gravity. It means that they can
be considered as the dark matter candidates.

\begin{acknowledgements}
The author is grateful to Prof. V.~S.~Barashenkov for references
on six-dimensional relativity in the first stage of this work.
\end{acknowledgements}

\appendix

\section{Notations and conventions \label{nots}}

The following notations are used throughout the paper.
Capital Latin letters $K,L,M,N$ represent indices
of the flat $M(3,3)$ space ($K \in \{1,2,3,4,5,6\}$)
with metric
\be
\eta_{\dbss{KL}} = \mbox{diag}\{-1,-1,-1,+1,+1,+1\}\,.
\label{dan1}
\ee
All the 6D indices are lowered and raised by the matrix
$\eta_{\dbss{KL}}$ and its inverse $\eta^{KL}=\eta_{\dbss{KL}}$.
Small Latin letters $a,b,c,d$ indicate indices of
the flat 4D space ($a \in \{1,2,3,4\}$) with metric
\be
\eta'_{ab} = \mbox{diag}\{-1,-1,-1,+1\}\,.
\label{dan3}
\ee
The following representation for the $4 \times 4$ Dirac matrices
is used:
\be
\gamma^k = \left( \begin{array}{lc}
\bf{0} & -\bfsigma^k \\ \bfsigma^k & \,\bf{0}
\end{array} \right),\quad
\gamma^4 = \left( \begin{array}{cc}
\bf{0} & \bf{1} \\ \bf{1} & \bf{0}
\end{array} \right),\quad
\gamma^5 = \left( \begin{array}{rr}
-\bf{1} & \bf{0} \\ \bf{0} & \bf{1}
\end{array} \right),
\label{dan5}
\ee
where $k=1,2,3$, and
boldface characters denote $2 \times 2$ matrices
including conventional Pauli matrices $\bfsigma^k$:
\be
\bfsigma^1 = \left( \begin{array}{cc}
0 & 1 \\ 1 & 0
\end{array} \right),\quad
\bfsigma^2 = \left( \begin{array}{rr}
0 & -i \\ i & 0
\end{array} \right),\quad
\bfsigma^3 = \left( \begin{array}{rr}
1 & 0 \\ 0 & -1
\end{array} \right).
\label{dan6}
\ee

\section{Properties of the determinants
 of the  $O(p,q)$\\ matrices \label{detprp}}

Let $C$ be a real matrix of the matrix representation of the
group $O(p,q)$, i.e. the following equality is fulfilled:
\be
C^{\,T}\eta\,C = \eta
\label{detp1}
\ee
where
\be
\eta = \left( \begin{array}{cc}
-I_p & 0 \\ 0 & I_q
\end{array} \right)\,,
\label{detp2}
\ee
$I_p$ and $I_q$ are the $p \times p$ and $q \times q$ identity
matrices. Let us write $C$ in the block form as
\be
C = \left( \begin{array}{cc}
A & b \\ a & B
\end{array} \right)
\label{detp3}
\ee
where $A$, $b$, $a$, and $B$ are the $p \times p$, $p \times q$,
$q \times p$, and $q \times q$ matrices, respectively.
Eqs. (\ref{detp1})--(\ref{detp3}) lead to the equalities:
\bea
&&A^TA = I_p + a^Ta\,,
\label{detp4}\\
&&B^TB = I_q + b^Tb\,,
\label{detp5}\\
&&a^TB = A^Tb\,.
\label{detp6}
\eea
The matrix $a^Ta$ is real, symmetric, and non-negative-definite.
Consequently, according to Eq.~(\ref{detp4}),
all the eigenvalues of the matrix $A^TA$ are equal to or greater than
unity. From this we obtain that $|\det(A)|\geqslant 1$.
Analogously one can prove that $|\det(B)|\geqslant 1$.

The following formula for the determinant of the matrix
of the form (\ref{detp3}) is known (see, e.g., \cite{G59}):
\be
\det(C) = \det(A)\det(B-aA^{-1}b)\,.
\label{detp7}
\ee
Using Eq.~(\ref{detp6}) we get $B^TaA^{-1}b=b^Tb$.
From this and from Eq.~(\ref{detp5}) it follows that
$B^T(B-aA^{-1}b)=B^TB-b^Tb=I_q$. Therefore
$\det(B-aA^{-1}b)=1/\det(B)$ and from Eq.~(\ref{detp7})
we find the final result:
$\det(C) = \det(A)/\det(B)$. Analogously one can obtain that
$\det(C) = \det(B)/\det(A)$. Because $|\det(C)|=1$ we also get
that $|\det(A)|=|\det(B)|$.

\section{Matrices of the spinor transformations
 and their basic properties \label{matprp}}

Let us introduce $4 \times 4$ matrices
$\alpha^K$ and $\tilde{\alpha}^K$
\bea
&&\alpha^A = \gamma^5\,\gamma^A\,,\quad \alpha^6 = i\,\gamma^5\,,
\label{mprp1}\\
&&\tilde{\alpha}^A = \gamma^A\,\gamma^5\,,
\quad \tilde{\alpha}^6 = -i\,\gamma^5
\label{mprp2}
\eea
where $A \in \{1,2,3,4,5\}$ and the representation of the Dirac
matrices given in Appendix~\ref{nots} is implied.
Using the above formulas, let us define the following matrices:
\bea
\beta^{KLM} &=& \frac{i}{2}\,
\left( \delta^{K}_{\dbss{K'}}\,\delta^{M}_{\dbss{M'}} -
\delta^{K}_{\dbss{M'}}\,\delta^{M}_{\dbss{K'}} \right)
\alpha^{K'}\,\tilde{\alpha}^L\,\alpha^{M'}\,,
\label{mprp3}\\
\tilde{\beta}^{KLM} &=& \frac{i}{2}\,
\left( \delta^{K}_{\dbss{K'}}\,\delta^{M}_{\dbss{M'}} -
\delta^{K}_{\dbss{M'}}\,\delta^{M}_{\dbss{K'}} \right)
\tilde{\alpha}^{K'}\,\alpha^L\,\tilde{\alpha}^{M'}\,,
\label{mprp4}
\eea
\bea
S^{KL} &=& \frac{i}{2}\,
\left( \eta^{KL} -
\tilde{\alpha}^L\,\alpha^{K} \right)\,,
\label{mprp5}\\
\tilde{S}^{KL} &=& \frac{i}{2}\,
\left(\alpha^{K}\,\tilde{\alpha}^L -
\eta^{KL} \right)\,.
\label{mprp6}
\eea

From these definitions and the properties of the Dirac matrices
it follows that:
\begin{itemize}
\item[(a)]
Matrices $\beta^{KLM}$, $\tilde{\beta}^{KLM}$,
$S^{KL}$, and $\tilde{S}^{KL}$ are totally antisymmetric
in the 6D indices $K,L,M$.
The matrices $\beta^{KLM}$ ($\tilde{\beta}^{KLM}$)
have properties of the anti-self-dual (self-dual) tensors
in the 6D indices, i.e.:
\bea
\beta^{KLM} &=& - \frac{1}{6}\,\ve^{KLMK'L'M'}
\beta_{\dbss{K'L'M'}}\,,
\label{asdp}\\
\tilde{\beta}^{KLM} &=& \hphantom{-}  \frac{1}{6}\,\ve^{KLMK'L'M'}
\tilde{\beta}_{\dbss{K'L'M'}}
\label{sdp}
\eea
where $\ve^{KLMK'L'M'}$ is the totally antisymmetric 6D tensor
with $\ve^{123456}=-1$.
There are only 10 independent matrices $\beta^{KLM}$ and
15 independent matrices $S^{KL}$. The same is true for
the matrices $\tilde{\beta}^{KLM}$ and $\tilde{S}^{KL}$.
In the explicit form we have:
\be
\alpha^a = - \tilde{\alpha}^a = \gamma^5\,\gamma^a\,,\quad
\alpha^5 = \tilde{\alpha}^5 = 1\,,\quad
\alpha^6 = - \tilde{\alpha}^6 = i\,\gamma^5\,,
\nonumber
\ee
\begin{align*}
\beta^{\,a56} & = \tilde{\beta}^{\,a56} = \gamma^a\,, &
\beta^{\,ab5} & = \tilde{\beta}^{\,ab5} = i\sigma^{ab}\,,\\
\beta^{\,ab6} & = - \tilde{\beta}^{\,ab6} =
\frac{i}{2}\,\ve^{abcd}\sigma_{\dbss{cd}}\,, &
\beta^{\,abc} & = - \tilde{\beta}^{\,abc} =
- \ve^{abcd}\gamma_{\dbss{d}}\,,
\end{align*}
\begin{align*}
S^{\,ab} & = \tilde{S}^{\,ab} = \frac{i}{2}\sigma^{ab}\,, &
S^{\,a5} & = - \tilde{S}^{\,a5} = - \frac{i}{2}\gamma^5\gamma^a\,,\\
S^{\,a6} & = \tilde{S}^{\,a6} = - \frac{1}{2}\gamma^a\,, &
S^{\,56} & = - \tilde{S}^{\,56} = - \frac{1}{2}\gamma^5
\end{align*}
where $\sigma^{ab}=\frac{1}{2}[\gamma^a,\gamma^b]$ and
$\ve^{abcd}$ is the totally antisymmetric 4D tensor
with $\ve^{1234}=-1$.
\item[(b)]
Matrices
$\{\alpha^K,\,\beta^{KLM}\}$,
$\{\tilde{\alpha}^K,\,\tilde{\beta}^{KLM}\}$,
$\{1,\,S^{KL}\}$, and $\{1,\,\tilde{S}^{KL}\}$
form four complete sets. This means that for any $4 \times 4$
complex matrices $A$ and $B$ the following equalities are
fulfilled:
\be
\mbox{Tr}(A\alpha^K)\,\mbox{Tr}(B\tilde{\alpha}_{\dbss{K}})
+\txts{\frac{1}{12}}\,\mbox{Tr}(A\beta^{KLM})\,
\mbox{Tr}(B\tilde{\beta}_{\dbss{KLM}})
=4\,\mbox{Tr}(AB)\,,
\label{mprp7}
\ee
\bea
\mbox{Tr}(A)\,\mbox{Tr}(B)
+2\,\mbox{Tr}(AS^{KL})\,\mbox{Tr}(BS_{\dbss{KL}})
&=&4\,\mbox{Tr}(AB)\,,
\label{mprp8}\\
\mbox{Tr}(A)\,\mbox{Tr}(B)
+2\,\mbox{Tr}(A\tilde{S}^{KL})\,\mbox{Tr}(B\tilde{S}_{\dbss{KL}})
&=&4\,\mbox{Tr}(AB)\,.
\label{mprp9}
\eea
\item[(c)]
Matrices $S^{KL}$ and $\tilde{S}^{KL}$ are the generators
of the Lie algebra $so(3,3)$, i.e., we have:
\be
\bigl[ S^{\,KL},S^{\,K'L'} \bigr] =
i \bigl(\,\eta^{\,KL'}S^{\,LK'}
+ \eta^{\,LK'}S^{\,KL'}
- \eta^{\,KK'}S^{\,LL'}
- \eta^{\,LL'}S^{\,KK'} \bigr)
\label{so33}
\ee
and the same for the matrices $\tilde{S}^{KL}$.
\end{itemize}

The last equation follows from the general formula for the
product of the $S^{KL}$ matrices
\bea
S^{\,KL}S^{\,K'L'} &=& \frac{1}{4}\,
\bigl(\,\eta^{\,KK'}\eta^{\,LL'} - \eta^{\,KL'}\eta^{\,LK'}
\bigr)
\nonumber\\
&+& \frac{i}{2}\,\bigl(\,\eta^{\,KL'}S^{\,LK'}
+ \eta^{\,LK'}S^{\,KL'}
- \eta^{\,KK'}S^{\,LL'}
- \eta^{\,LL'}S^{\,KK'} \bigr)
\nonumber\\
&+& \frac{i}{4}\,\ve^{KLK'L'MN}S_{\lindb{MN}}
\label{sprod}
\eea
which can be obtained using the algebra of the Dirac matrices.
The formula for the product of the $\tilde{S}^{KL}$ matrices
has the same form but with the opposite sign in front of the
last term.
Two more important relations also follow from the algebra of the
Dirac matrices, namely
\bea
&&\alpha^K S^{LM} - \tilde{S}^{LM} \alpha^K
= i\,(\eta^{KL} \alpha^M - \eta^{KM} \alpha^L)\,,
\label{mprp21}\\
&&\tilde{\alpha}^K \tilde{S}^{LM} - S^{LM} \tilde{\alpha}^K
= i\,(\eta^{KL} \tilde{\alpha}^M - \eta^{KM} \tilde{\alpha}^L)\,.
\label{mpr21b}
\eea

Consider the symmetry properties of the $\alpha$, $\beta$, and $S$
matrices. Let us define for an arbitrary $4 \times 4$ matrix $A$
the matrix
$A^{\langle T \rangle}$ as
\be
A^{\langle T \rangle} = - \gamma^4 \gamma^2 A^T \gamma^2 \gamma^4\,.
\label{mprp10}
\ee
It is easy to check that the following relations
are fulfilled:
\be
(A^{\langle T \rangle})^{\langle T \rangle} = A\,,\quad
(A B)^{\langle T \rangle} =
B^{\langle T \rangle} A^{\langle T \rangle}\,,
\label{mprp11}
\ee
\be
(\alpha^K)^{\langle T \rangle} = \alpha^K\,,\quad
(\tilde{\alpha}^K)^{\langle T \rangle} =
\tilde{\alpha}^K\,,
\label{mprp12}
\ee
\be
(\beta^{KLM})^{\langle T \rangle} = - \beta^{KLM}\,,\quad
(\tilde{\beta}^{KLM})^{\langle T \rangle} = - \tilde{\beta}^{KLM}\,,
\label{mprp13}
\ee
\be
(S^{KL})^{\langle T \rangle} = - \tilde{S}^{KL}\,,\quad
(\tilde{S}^{KL})^{\langle T \rangle} = - S^{KL}\,.
\label{mprp14}
\ee
Complex conjugation leads to the equalities:
\be
\gamma^2\,(\alpha^{K})^* \gamma^2 = - \alpha^K\,,\quad
\gamma^2\,(\tilde{\alpha}^K)^* \gamma^2 =
- \tilde{\alpha}^K\,,
\label{mprp15}
\ee
\be
\gamma^2\,(\beta^{KLM})^* \gamma^2 = \beta^{KLM}\,,\quad
\gamma^2\,(\tilde{\beta}^{KLM})^* \gamma^2 = \tilde{\beta}^{KLM}\,,
\label{mprp16}
\ee
\be
\gamma^2\,(S^{KL})^* \gamma^2 = S^{KL}\,,\quad
\gamma^2\,(\tilde{S}^{KL})^* \gamma^2 = \tilde{S}^{KL}\,.
\label{mprp17}
\ee
From Eqs. (\ref{mprp10}), (\ref{mprp12})--(\ref{mprp17})
it follows that
\be
\gamma^4\,(\alpha^K)^{\dag} \gamma^4 = \alpha^K\,,\quad
\gamma^4\,(\tilde{\alpha}^K)^{\dag} \gamma^4 =
\tilde{\alpha}^K\,,
\label{mprp18}
\ee
\be
\gamma^4\,(\beta^{KLM})^{\dag} \gamma^4 = \beta^{KLM}\,,\quad
\gamma^4\,(\tilde{\beta}^{KLM})^{\dag} \gamma^4
= \tilde{\beta}^{KLM}\,,
\label{mprp19}
\ee
\be
\gamma^4 (S^{KL})^{\dag} \gamma^4 = \tilde{S}^{KL}\,,\quad
\gamma^4 (\tilde{S}^{KL})^{\dag} \gamma^4 = S^{KL}\,.
\label{mprp20}
\ee
On the other hand,
using Eqs. (\ref{mprp1})--(\ref{mprp6}) one can show that
the following equalities are fulfilled:
\be
\gamma^4\,\alpha^K \gamma^4 =
{\cal{P}}^K_{\linda{K'}}\,\tilde{\alpha}^{K'}\,,\quad
\gamma^4\,\tilde{\alpha}^K \gamma^4 =
{\cal{P}}^K_{\linda{K'}}\,\alpha^{K'}\,,
\label{sref1}
\ee
\be
\gamma^4\,\beta^{KLM} \gamma^4 =
{\cal{P}}^K_{\linda{K'}}\,{\cal{P}}^L_{\linda{L'}}\,
{\cal{P}}^M_{\linda{M'}}\,\tilde{\beta}^{K'L'M'}\,,\quad
\gamma^4\,\tilde{\beta}^{KLM} \gamma^4 =
{\cal{P}}^K_{\linda{K'}}\,{\cal{P}}^L_{\linda{L'}}\,
{\cal{P}}^M_{\linda{M'}}\,\beta^{K'L'M'}\,,
\label{sref2}
\ee
\be
\gamma^4\,S^{KL} \gamma^4 =
{\cal{P}}^K_{\linda{K'}}\,{\cal{P}}^L_{\linda{L'}}\,
\tilde{S}^{K'L'}\,,\quad
\gamma^4\,\tilde{S}^{KL} \gamma^4 =
{\cal{P}}^K_{\linda{K'}}\,{\cal{P}}^L_{\linda{L'}}\,
S^{K'L'}
\label{sref3}
\ee
where the matrix of the space reflection ${\cal{P}}$
is defined by Eq.~(\ref{elg6}).

Finally, for the traces we have
\be
\mbox{Tr}\,(S^{KL}) =
\mbox{Tr}\,(\tilde{S}^{KL}) = 0\,,
\label{mprp22}
\ee
\be
\mbox{Tr}\,(\beta^{KLM} \tilde{\alpha}^N) =
\mbox{Tr}\,(\tilde{\beta}^{KLM} \alpha^N) = 0\,,
\label{mprp23}
\ee
\be
\mbox{Tr}\,(S^{KL} S^{MN}) =
\mbox{Tr}\,(\tilde{S}^{KL} \tilde{S}^{MN}) =
\eta^{KM}\eta^{LN} - \eta^{KN}\eta^{LM}\,,
\label{mprp25}
\ee
\be
\mbox{Tr}\,(\alpha^K \tilde{\alpha}^L)
= 4\,\eta^{KL}\,.
\label{mprp24}
\ee

\section{Proof of the equations connecting $GL(4,M)$ and $EL(3,3)$
 groups \label{prfeqs}}

Let us prove Eqs. (\ref{ms5}) and (\ref{ms6}).
First of all, notice that in accordance with
Eqs. (\ref{ms2}), (\ref{ms4}), and (\ref{mprp10}) we have
$S^{\langle T \rangle} = \barst$ if $S \in GL(4,M)$.
Therefore [see Eqs. (\ref{mprp11}) and (\ref{mprp12})]:
\be
(\barst\,\alpha^K S)^{\langle T \rangle} = \barst\,\alpha^K S\,.
\label{ms7}
\ee
Expanding the matrix $\barst\,\alpha^K S$ in the complete set
$\{\alpha^K,\,\beta^{KLM}\}$, we obtain Eq.~(\ref{ms5})
because the terms containing $\beta^{KLM}$ vanish due to
the opposite signs in the symmetry relations
(\ref{mprp12}), (\ref{mprp13}), and (\ref{ms7}).
Expression~(\ref{ms6}) for the matrix $\Lambda$
follows from Eqs. (\ref{ms5}) and (\ref{mprp24}).
Finally, reality of $\Lambda$ follows from
Eqs. (\ref{ms2}) and (\ref{mprp15}).
In the same way we obtain Eqs. (\ref{ms8}) and (\ref{ms9}).

Consider the properties of the matrices $\Lambda$ and
$\tilde{\Lambda}$ defined by Eqs. (\ref{ms6}) and (\ref{ms9}).
First of all, the matrices $\Lambda$ form a group.
To see this, let us take the matrix $\Lambda$
in form (\ref{ms6}) and the matrix $\Lambda'$ in the form:
\be
\Lambda'\vphantom{\Lambda}^K_{\lindb{L}} = \frac{1}{4}\,
\mbox{Tr}\left( \barst\vps^{\,\prime}\,\alpha^K S^{\,\prime}\,
\tilde{\alpha}_{\dbss{L}} \right)\,.
\label{ms10}
\ee
Using the completeness
relation (\ref{mprp7}), Eqs. (\ref{ms5}),
(\ref{ms6}), (\ref{mprp23}), and (\ref{ms10})
we obtain:
\be
\Lambda^K_{\lindb{L}}
\Lambda'\vphantom{\Lambda}^L_{\lindb{M}} =
\Lambda''\vphantom{\Lambda}^K_{\lindb{M}} =
\frac{1}{4}\,\mbox{Tr}
\left( \barst\vps^{\,\prime\prime}\,\alpha^K S^{\,\prime\prime}\,
\tilde{\alpha}_{\dbss{M}} \right)
\label{ms11}
\ee
where $S^{\,\prime\prime} = S S^{\,\prime}$.
Analogously, one can show that the matrices $\tilde{\Lambda}$
also form a group.
In the same way, but with the use of
Eqs. (\ref{ms8}), (\ref{ms9}), and (\ref{mprp24}) we get
\be
\Lambda^K_{\lindb{L}}\,\tilde{\Lambda}^{L}_{\lindb{M}}=
\tilde{\Lambda}^K_{\lindb{L}}\,\Lambda^{L}_{\lindb{M}}=
\delta^{K}_{\dbss{M}}\,.
\label{ms12}
\ee
Thus, $\tilde{\Lambda} = \Lambda^{-1}$.

Let us prove Eq.~(\ref{ms13}).
At first, consider the case $\det(S)=+1$.
It is known (see, e.g., \cite{G59}) that any real matrix $R$
with $\det(R)=1$ can be represented in the form:
$R=O_{\linda{1}} D\,O_{\linda{2}}$ where the matrices
$O_{\linda{1}}$, $D$, and $O_{\linda{2}}$ are real,
$O_{\linda{1}}$ and $O_{\linda{2}}$ are orthogonal,
$D$ is diagonal and positive-definite, and
$\det(O_{\linda{1}})=\det(O_{\linda{2}})=\det(D)=1$.
Any real orthogonal matrix $O$ with $\det(O)=1$ can be
represented in the form: $O=e^A$ where $A=A^*=-A^T$.
From this it follows that $R$ can be represented in the form:
$R=e^{A}e^{A'}e^{A''}$ where the matrices $A$, $A'$, and $A''$
are real and traceless.
Using this result and the isomorphism
between $GL(4,M)$ and $GL(4,R)$, we obtain that any matrix
$S \in SL(4,M)$ [i.e., any $S \in GL(4,M)$ with $\det(S)=1$]
can be represented in the form
\be
S = e^{\Theta}\,e^{\Theta'} e^{\Theta''}
\label{ms14}
\ee
where
\be
\gamma^2\,\Theta^* \gamma^2 = -\Theta\,,\quad
\mbox{Tr}(\Theta)=0\,,
\label{ms15}
\ee
and the same conditions are fulfilled for $\Theta'$ and $\Theta''$.
Thus, by virtue of the group properties, it is sufficient to prove
Eq.~(\ref{ms13}) for $S = e^{\Theta}$.

From the completeness of the set $\{1,\,S^{KL}\}$
and from Eqs. (\ref{mprp17}) and (\ref{mprp22})
it follows that any $4 \times 4$ complex matrix $\Theta$
satisfying conditions (\ref{ms15}) can be represented in the form
\be
\Theta = i\,\theta^{L}_{\lindb{K}}S^{K}_{\lindb{L}}
\label{ms16}
\ee
where $S^{K}_{\lindb{L}}=S^{KL'}\eta_{\dbss{L'L}}$ and
the matrix $\theta^{L}_{\lindb{K}}$ is real and possesses the property
\be
\theta^{L}_{\lindb{K}} = -\eta^{LL'}\eta_{\dbss{KK'}}\,
\theta^{K'}_{\lindb{L'}}\,.
\label{ms17}
\ee
In shorthand notation we have:
\be
\eta\,\theta\,\eta=-\theta^{\,T}\,.
\label{ms18}
\ee
Using Eqs. (\ref{mprp21}), (\ref{ms16}), and (\ref{ms17})
we obtain the following equality:
\be
\alpha^{K}e^{\Theta} = e^{-\bar{\Theta}}
\Lambda^K_{\lindb{L}}(\theta)\,\alpha^{L}
\label{ms19}
\ee
where $\bar{\Theta}=\gamma^4\,\Theta^{\dag}\gamma^4$,
$\,\Lambda(\theta)=e^{\,2\theta}$.
Eq.~(\ref{ms19}) is a particular case of Eq.~(\ref{ms5})
for $S = e^{\Theta}$, so Eq.~(\ref{ms6}) is also true.
Further, with the help of Eq.~(\ref{ms18}) we get:
$\eta\,\Lambda(\theta)\,\eta=\Lambda^T(-\theta)$.
On the other hand, it is obviously that
$\Lambda(-\theta)=\Lambda^{-1}(\theta)$.
Therefore, we have:
$\Lambda^T(\theta)\,\eta\,\Lambda(\theta)=\eta$.
This proves Eq.~(\ref{ms13}) for the case $\det(S)=+1$.

Let $\det(S)=-1$. Any matrix $S_- \in GL(4,M)$ with $\det(S_-)=-1$
can be represented in the form $S_-=S_+S_P$ where
$S_+ \in GL(4,M)$, $\,\det(S_+)=+1$,
\be
S_P = \frac{1}{2}\, (1 -i\,\gamma^1 -\gamma^2\gamma^5
- \gamma^3\gamma^4)\,.
\label{ms20}
\ee
The matrix $S_P \in GL(4,M)$. It has $\det(S_P)=-1$ and generates
the space-time permutation. Namely, substituting $S_P$ into
Eq.~(\ref{ms6}) we obtain $\Lambda=\bar{I}_6$ with
the matrix $\bar{I}_6$ defined in Eq.~(\ref{elg7}). Since
$\,\bar{I}^{\,T}_6\,\eta\,\bar{I}^{\vphantom{2}}_6 = - \eta$,
the matrix $\Lambda$ generated by $S_-$ also satisfies
Eq.~(\ref{ms13}).

The extension of the proof of Eq.~(\ref{ms13}) to the general case
is obvious because any matrix $S \in GL(4,M)$ can be represented in
one of the two forms, namely: $S=\lambda S_+$ or $S=\lambda S_-$ where
$S_{\pm} \in GL(4,M)$, $\,\det(S_{\pm})=\pm 1$, $\,\lambda$
is a real number. Eq.~(\ref{ms21}) is proved analogously.

The following remark is related to the property of the mapping
$SL(4,M) \to L^{\uparrow}_+(3,3)$. Consider matrices of the form:
$\Lambda_{\linda{+}}=e^{\,2\theta}e^{\,2\theta'}e^{\,2\theta''}$
where the $6 \times 6$ real matrices $\theta$, $\theta'$, and
$\theta''$ satisfy Eq.~(\ref{ms18}). From the above derivation
it follows that the matrices $\Lambda_{\linda{+}}$ form a group
$L_{\linda{\Lambda +}}$, and $SL(4,M)$ is mapped {\it onto}
$L_{\linda{\Lambda +}}$.
Obviously, the group $L_{\linda{\Lambda +}}$ is connected and
$L_{\linda{\Lambda +}} \subseteq L^{\uparrow}_+(3,3)$.
On the other hand, it is easy to see that
$\Lambda L_{\linda{\Lambda +}} \Lambda^{-1} \subseteq
L_{\linda{\Lambda +}}$ for all $\Lambda$ in $L^{\uparrow}_+(3,3)$.
Since the group $L^{\uparrow}_+(3,3)$ is simple, we obtain that
$L_{\linda{\Lambda +}} = L^{\uparrow}_+(3,3)$.

%

%
\end{document}